\documentclass[twocolumn]{aastex62}
\usepackage[utf8]{inputenc}
\DeclareUnicodeCharacter{2212}{-}
\usepackage{graphicx}
\graphicspath{{./}{Figures}}
\usepackage{amsmath}
\usepackage{ulem}
\usepackage{breqn}
\usepackage{afterpage}
\usepackage{enumitem}
\usepackage{xcolor}
\usepackage{float}
\usepackage{xspace} 
\usepackage{url}
\usepackage{fontawesome}
\setenumerate{itemsep=0mm}
\usepackage{lineno}
\usepackage{wrapfig}
\usepackage{placeins}

\newcommand{\sfive}{\ensuremath{S^5}\xspace}

\newcommand{\mstar}{\ensuremath{M_\star}\xspace}
\newcommand{\mvir}{\ensuremath{M_{\rm h}}\xspace}
\newcommand{\msun}{\ensuremath{M_\odot}\xspace}

\newcommand{\teff}{\ensuremath{{T_{\rm eff}}}\xspace}
\newcommand{\logg}{\ensuremath{{\log g}}\xspace}

\newcommand{\vhel}{\ensuremath{{v_{\rm hel}}}\xspace}

\newcommand{\dhelio}{\ensuremath{{d_{\rm hel}}}\xspace}

\newcommand{\vgsr}{\ensuremath{v_{\rm gsr}}\xspace}

\newcommand{\rh}{\ensuremath{{R_{1/2}}}\xspace}
\newcommand{\sigv}{\ensuremath{{\sigma_v}}\xspace}

\usepackage{amsmath}
\usepackage{amssymb}
\usepackage{xspace}
\usepackage{xifthen}
\usepackage{eso-pic}

\definecolor{forestgreen}{HTML}{228B22}
\definecolor{urlblue}{HTML}{000000}

\mathchardef\mhyphen="2D

\newlength{\dhatheight}

\newcommand{\unit}[1]{\ensuremath{\mathrm{\,#1}}\xspace}

\newcommand{\km}{\unit{km}}
\newcommand{\kms}{\km \second^{-1}}

\newcommand{\second}{\unit{s}}

\newcommand{\bandvar}[2][]{%
  \ifthenelse{\isempty{#1}}{\var{#2}}{\var{#2\_#1}}%
}

\newcommand{\feh}{{\ensuremath{\rm [Fe/H]}}\xspace}

\newcommand{\var}[1]{\ensuremath{\texttt{\MakeUppercase{#1}}}\xspace}

\providecommand\physrep{\ref@jnl{Phys.~Rep.}}%
\providecommand\apjs{\ref@jnl{ApJS}}%
\providecommand{\jcap}{\ref@jnl{JCAP}}%

\shorttitle{Tidal Disruption of Crater~2}
\shortauthors{$S^5$ Collaboration}

\begin{document}


\title{\textbf{\sfive: Tidal Disruption in Crater~2 and Formation of Diffuse Dwarf Galaxies in the Local Group} 
}


\newcommand{\MITPhysics}{Department of Physics, Massachusetts Institute of Technology, 77 Massachusetts Avenue, Cambridge, MA 02139, USA}
\newcommand{\MITKavli}{Kavli Institute for Astrophysics and Space Research, Massachusetts Institute of Technology, 77 Massachusetts Avenue, Cambridge, MA 02139, USA}
\newcommand{\UChicagoAA}{Department of Astronomy \& Astrophysics, University of Chicago, 5640 S. Ellis Avenue, Chicago, IL 60637, USA}
\newcommand{\KICP}{Kavli Institute for Cosmological Physics, University of Chicago, 5640 S. Ellis Avenue, Chicago, IL 60637, USA}
\newcommand{\IAGUSP}{Universidade de S\~ao Paulo, Instituto de Astronomia, Geof\'isica e Ci\^encias Atmosf\'ericas, Departamento de Astronomia, SP 05508-090, S\~ao Paulo, Brazil}
\newcommand{\NOIRLab}{NSF NOIRLab, Tucson, AZ 85719, USA}
\newcommand{\JINA}{Joint Institute for Nuclear Astrophysics--Center for the Evolution of the Elements (JINA-CEE), USA}
\newcommand{\SkAI}{NSF-Simons AI Institute for the Sky (SkAI), 172 E. Chestnut St., Chicago, IL 60611, USA}


\correspondingauthor{Guilherme Limberg}
\email{limberg@uchicago.edu}

\author[0000-0002-9269-8287]{Guilherme~Limberg}
\affiliation{\KICP}
\affiliation{\UChicagoAA}

\author[0000-0002-4863-8842]{Alexander~P.~Ji}
\affiliation{\UChicagoAA}
\affiliation{\KICP}
\affiliation{\SkAI}

\author[0000-0002-9110-6163]{Ting~S.~Li}
\affiliation{Department of Astronomy and Astrophysics, University of Toronto, 50 St. George Street, Toronto, ON M5S 3H4, Canada}
\affiliation{Dunlap Institute for Astronomy \& Astrophysics, University of Toronto, 50 St. George Street, Toronto, ON M5S 3H4, Canada}
\affiliation{Data Sciences Institute, University of Toronto, 17th Floor, Ontario Power Building, 700 University Avenue, Toronto, ON M5G 1Z5, Canada}

\author[0000-0002-8448-5505]{Denis~Erkal}
\affiliation{School of Mathematics and Physics, University of Surrey, Guildford GU2 7XH, UK}

\author[0000-0003-2644-135X]{Sergey~E.~Koposov}
\affiliation{Institute for Astronomy, University of Edinburgh, Royal Observatory, Blackford Hill, Edinburgh EH9 3HJ, UK}
\affiliation{Institute of Astronomy, University of Cambridge, Madingley Road, Cambridge CB3 0HA, UK}


\author[0000-0002-6021-8760]{Andrew~B.~Pace}
\thanks{Galaxy Evolution and Cosmology (GECO) Fellow}
\affiliation{Department of Astronomy, University of Virginia, 530 McCormick Road, Charlottesville, VA 22904, USA}


\author[0009-0005-5355-5899]{Andrew~P.~Li}
\affiliation{Dunlap Institute for Astronomy \& Astrophysics, University of Toronto, 50 St. George Street, Toronto, ON M5S 3H4, Canada}

\author[0000-0002-0428-849X]{Petra~Awad}
\affiliation{Leiden Observatory, Leiden University, PO Box 9513, NL-2300 RA Leiden, The Netherlands}

\author[0009-0004-5519-0929]{Alexandra~Senkevich}
\affiliation{School of Mathematics and Physics, University of Surrey, Guildford GU2 7XH, UK}

\author[0000-0001-7516-4016]{Joss~Bland-Hawthorn}
\affiliation{Sydney Institute for Astronomy, School of Physics, A28, The University of Sydney, NSW 2006, Australia}

\author[0000-0001-8536-0547]{Lara~Cullinane}
\affiliation{Leibniz-Institut f{\"u}r Astrophysik Potsdam (AIP), An der Sternwarte 16, D-14482 Potsdam, Germany}

\author[0000-0001-7019-649X]{Gary~Da~Costa}
\affiliation{Research School of Astronomy and Astrophysics, Australian National University, Canberra, ACT 2611, Australia}

\author[0000-0001-8251-933X]{Alex~Drlica-Wagner}
\affiliation{\UChicagoAA}
\affiliation{\KICP}
\affiliation{Fermi National Accelerator Laboratory, P.O. Box 500, Batavia, IL 60510, USA}
\affiliation{\SkAI}

\author{Rapha\"el~Errani}
\affiliation{McWilliams Center for Cosmology and Astrophysics, Department of Physics, Carnegie Mellon University, 5000 Forbes Avenue, Pittsburgh, PA 15213, USA}

\author[0000-0001-6957-1627]{Peter~S.~Ferguson}
\affiliation{DIRAC Institute, Department of Astronomy, University of Washington, 3910 15th Ave NE, Seattle, WA, 98195, USA}

\author[0000-0003-0120-0808]{Kyler~Kuehn}
\affiliation{Lowell Observatory, 1400 W Mars Hill Rd, Flagstaff,  AZ 86001, USA}

\author[0000-0003-3081-9319]{Geraint~F.~Lewis}
\affiliation{Sydney Institute for Astronomy, School of Physics, A28, The University of Sydney, NSW 2006, Australia}

\author[0000-0002-3430-4163]{Sarah~L.~Martell}
\affiliation{School of Physics, University of New South Wales, Sydney, NSW 2052, Australia}

\author{Jorge~Pe\~narrubia}
\affiliation{Institute for Astronomy, University of Edinburgh, Royal Observatory, Blackford Hill, Edinburgh EH9 3HJ, UK}

\author[0000-0003-2497-091X]{Nora~Shipp}
\affiliation{DIRAC Institute, Department of Astronomy, University of Washington, 3910 15th Ave NE, Seattle, WA, 98195, USA}

\author[0000-0001-7609-1947]{Yong~Yang}
\affiliation{Sydney Institute for Astronomy, School of Physics, A28, The University of Sydney, NSW 2006, Australia}

\author[0000-0003-1124-8477]{Daniel~B.~Zucker}
\affiliation{School of Mathematical and Physical Sciences, Macquarie University, Sydney, NSW 2109, Australia}
\affiliation{Macquarie University Research Centre for Astrophysics and Space Technologies, Sydney, NSW 2109, Australia}

\collaboration{($S^5$ Collaboration)}
\noaffiliation


\begin{abstract}

We present results of a spectroscopic campaign around the diffuse dwarf galaxy Crater~2 (Cra2) and its tidal tails as part of the Southern Stellar Stream Spectroscopic Survey (\sfive). Cra2 is a Milky Way dwarf spheroidal 
satellite with extremely cold kinematics, but a huge size similar to the Small Magellanic Cloud, which may be difficult to explain within collisionless cold dark matter. We identify 143 Cra2 members, of which 114 belong to the galaxy's main body and 29 are deemed part of its stellar stream. We confirm that Cra2 is dynamically cold (central velocity dispersion $2.51^{+0.33}_{-0.30}\,\kms$) and also discover a $\approx$7$\sigma$ velocity gradient consistent with its tidal debris track. We separately estimate the stream's internal velocity dispersion to be $5.74^{+0.98}_{-0.83}\,\kms$. We develop a suite of $N$-body simulations with both cuspy and cored density profiles on a realistic Cra2 orbit 
to compare with \sfive observations. We find that the velocity dispersion ratio between Cra2 stream and galaxy ($2.30^{+0.41}_{-0.35}$)
is difficult to reconcile with a cuspy halo with fiducial concentration and an initial mass predicted by standard stellar mass--halo mass relationships. Instead, either a cored halo with relatively small core radius or a low-concentration cuspy model can reproduce this ratio. Despite tidal mass loss, Cra2 is metal-poor ($\langle \rm[Fe/H]\rangle=-2.16\pm0.04$) compared to the stellar mass--metallicity relation for its luminosity. Other diffuse dwarf galaxies similar to Cra2 in the Local Group (Antlia~2 and Andromeda~19) also challenge galaxy formation models. Finally, we discuss possible formation scenarios for Cra2, including ram-pressure stripping of a gas-rich progenitor combined with tides.
\end{abstract}


\vspace{-1.6cm}

\section{Introduction} \label{sec:intro}
\setcounter{footnote}{21}

Dwarf galaxies occupy the extreme low-mass end of the galaxy luminosity function \citep[e.g.,][]{koposov2008_LF, Drlica-Wagner2020census, chinyitan2025centus}. These are the most chemically pristine and dark-matter dominated systems in the Universe \citep{Tolstoy2009, simon2019}. Their stellar populations provide insights on the drivers of galaxy evolution at small scales \citep{Sales2022rev} and their internal dynamics are sensitive to the physics of dark matter \citep{BattagliaNipoti2022rev}. Therefore, understanding the assembly of dwarf galaxies is critical to achieve a complete understanding of the multi-scale process of galaxy formation through the interplay between baryons and their hosting dark matter halos \citep{wechslerTinker2018}.

\begin{figure*}[pt!]
\centering
\includegraphics[width=2.1\columnwidth]{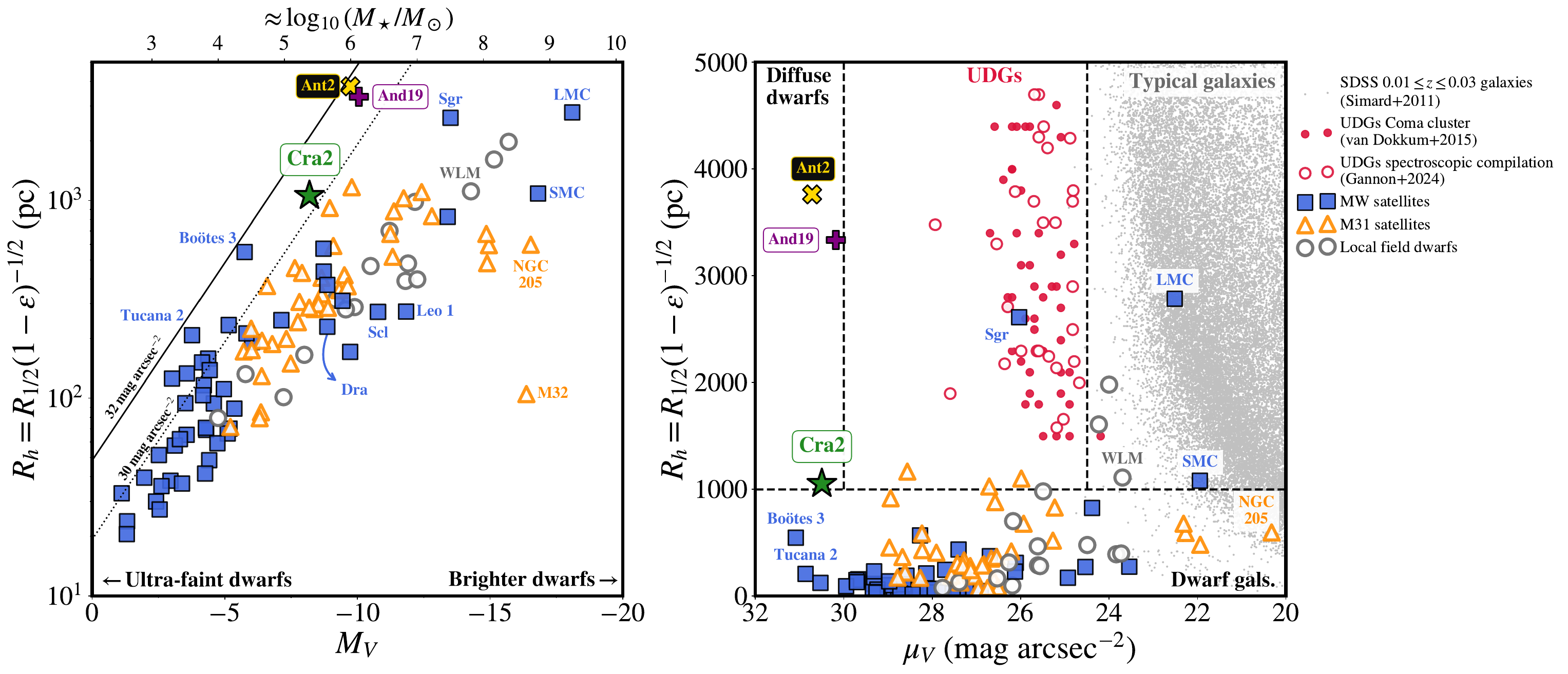}
\caption{Left: absolute $V$-band magnitude ($M_V$) vs. size diagram for Local Group dwarf galaxies \citep[compilation by][]{pace2025lvdb}. The vertical axis corresponds to the half-light major axis $R_h = \rh(1-\epsilon)^{-1/2}$, where $\rh$ is the circularized half-light radius and $\epsilon$ is the ellipticity. The corresponding stellar masses ($M_\star$) are computed as in Equation \ref{eq:mstar}. Diagonal lines exhibit constant values of surface brightness following annotations. Diffuse dwarfs Crater~2 (green `$\star$' symbol), Antlia~2 (yellow `X'), and Andromeda~19 (purple `$+$') are highlighted. Milky Way satellites, M31 satellites, and Local Group field dwarf galaxies are plotted as blue squares, orange triangles, and gray circles, respectively. Right: $V$-band surface brightness averaged within $R_h$ \citep[$\mu_V$; as in][]{gannon2024udgs} vs. size ($R_h$). Ultra-diffuse galaxies (UDGs) in the Coma cluster \citep{vanDokkum2015udgs} are shown as the red dots. White circles with red edges represent other UDGs from the \citet{gannon2024udgs}spectroscopic compilation. Gray dots are typical galaxies from the Sloan Digital Sky Survey data release 7 within the redshift range $0.01 \leq z \leq 0.03$ \citep[][and references therein]{simard2011sdss}.
\label{fig:rhalf_mu}}
\end{figure*}

Crater~2 (Cra2) is a Milky Way (MW) dwarf spheroidal (dSph) satellite 
and one of the most diffuse galaxies ever identified \citep{Torrealba2016cra2}. This faint dwarf galaxy has a size (circularized half-light radius; $\rh$) similar to the Small Magellanic Cloud \citep[SMC;][$\sim$1\,kpc]{munoz2018struc}, but $\sim$1000$\times$ lower stellar mass\footnote{Assuming the mean stellar mass-to-light ratio values from \citet{woo2008masstolightratios} for star-forming ($1.0\,\msun/L_\odot$) and quenched ($1.6\,\msun/L_\odot$) dwarf galaxies for SMC and Cra2, respectively.}. Its huge size and cold line-of-sight velocity dispersion \citep[$\sigv \sim 2.5\,\kms$;][]{Caldwell2017cra2, ji2021_cra2_ant2} make Cra2 one of the most severe outliers in the scaling relation between \sigv and \rh, i.e., a proxy for the mass--size relation \citep[e.g.,][and see \citealt{errani2022} for a recent version]{walker2010_mass_size_relat}, in the Local Group (see left panel of Figure \ref{fig:rhalf_mu}). The old stellar population \citep[$>$10\,Gyr;][]{walker2019cra2} and lower-than-expected metallicity for its luminosity \citep{ji2021_cra2_ant2} add to the remarkable properties of Cra2.

Galaxy evolution within collisionless cold dark matter (CDM) does offer a natural way to enlarge galaxies via tidal disruption \citep{Penarrubia2008tides, Errani2015tidalEvo}. However, even with tides, dynamical $N$-body simulations of `cuspy' Navarro-Frenk-White (NFW) halos \citep{NFW1996halos, nfw1997} struggle to reconcile Cra2's large \rh with its tiny \sigv \citep{sanders2018cra2, Borukhovetskaya2022cra2, errani2022}. A possible solution is to invoke that Cra2 is embedded in an NFW halo with either an unusually low concentration \citep[see][]{amorisco2019cra2} or extremely low halo mass for its luminosity. Another possibility is that Cra2 inhabits a `cored' halo instead, with a constant central density \citep[e.g.,][]{deBlok2010coreCusp}, which would allow for extra enlargement and suppression of \sigv for a fixed amount of mass loss \citep{fu2019cra2}. The issue with this hypothesis is that Cra2's stellar mass ($\mstar \sim 3\times10^5\,\msun$) is not large enough for energetic supernovae feedback to destroy its central cusp and create a core \citep[see review by][]{Bullock2017review}. 

As an alternative to a cored dark matter halo, results from hydrodynamical cosmological simulations imply that extreme \mstar loss, perhaps $\sim$99\%, could reproduce Cra2's structural properties even in the normal-concentration NFW case \citep{Fattahi2018satellite_sims, applebaum2021DC_justLeague_sims}. However, since Cra2 is metal-poor compared to the Local Group's stellar mass--metallicity relation \citep[MZR;][]{Kirby2013, simon2019}, this argument has been mostly rejected since a tidally disrupted galaxy should appear to be metal-rich instead \citep[][and see \citealt{riley2025aurigaMZR}]{ji2021_cra2_ant2}, as is the case for Sagittarius dSph \citep[e.g.,][]{Hayes2020}.

Given the lack of a satisfactory mechanism to fabricate galaxies like Cra2, we face the exciting possibility that a modification to the CDM model might be required. For example, self-interacting dark matter, where (postulated) dark matter particles can scatter and thermalize \citep{spergel2000sidm}, might be able to explain Cra2's structural properties, including both size and velocity dispersion \citep{Kahlhoefer2019sidmMWsats, lovell2023DMmodels, Zhang2024sidmCra2}. Tidal stripping with fuzzy (or wave) dark matter, where the quantum uncertainty principle itself creates a constant density core in dwarf galaxies \citep{WayneHu2000fuzzyDM, Schive2014waveDMsims}, might also reproduce Cra2's extended morphology and cold kinematics \citep{Pozo2022waveDMcrater2}. We note, however, that this scenario has already been strongly constrained by dwarf kinematics itself \citep{DalalKravtsov2022fuzzyDM}. Lastly, Modified Newtonian Dynamics \citep{milgrom1983mond} correctly predicts Cra2's \sigv value to be between $\sim$1.5\,$\kms$ and 3.0\,$\kms$ \citep{McGaugh2016mondCra2}, although such model is unable to reproduce large-scale structure formation.

%
Regardless of its extraordinary properties, Cra2 is not completely unique in the Local Group. Both the MW satellite Antlia~2 \citep[Ant2;][]{Torrealba2019ant2} and M31 satellite Andromeda~19 \citep[And19;][]{McConnachie2008andXIX} display comparable properties, such as extended morphology, cold kinematics, old stellar populations, and low metallicity \citep[e.g.,][]{Collins2020andXIX, collins2022andXIX, ji2021_cra2_ant2}. These nearby diffuse dwarfs resemble external ``ultra-diffuse'' galaxies \citep[UDGs;][]{vanDokkum2015udgs, koda2015udgs} usually found in dense cluster environments \citep{vanderBurg2017udgs}. We refer the interested reader to several recent works about UDG properties and their stellar populations in galaxy clusters \citep{buzzo2022udgs, Buzzo2024udgs, gannon2022udgs, LaMarca2022udgs, Ferre-Mateu2023udgs, iodice2023udgs, Marleau2024udgs, buttitta2025udgs, doll2026udgs}. Despite similarities, Cra2 and its siblings are much more extreme and lower-mass systems than UDGs; these local dwarfs are $\gtrsim$100$\times$ fainter than the prototypical UDG population (right panel of Figure \ref{fig:rhalf_mu}). Nevertheless, UDGs raise similar concerns, and solutions, regarding their possible formation pathway(s) \citep{DiCintio2017udgs, carleton2019udgs, ManceraPina2019udgs, Kong2022udgs, nadler2023sidmUDGs, benavides2023udgs}.

The leading hypotheses, either within or outside CDM, for the formation of Cra2, and its analogs Ant2 and And19, require tidal disruption. \citet{coppi2024cra2} recently detected a linear overdensity of variable RR Lyrae (RRL) stars around Cra2. This stellar stream aligns with the on-sky proper motion vector of Cra2 and strongly suggests that this dwarf galaxy is indeed experiencing heavy tidal mass loss. In this contribution, we investigate the chemodynamical properties of Cra2 and its stellar stream with red giant-branch (RGB) stars observed as part of the Southern Stellar Stream Spectroscopic Survey (\sfive) program \citep{Li2019s5}. We explore the kinematics of the Cra2 stellar stream to infer properties of its dark matter density profile. The chemistry along the Cra2 stream will also inform us, for
example, if the most metal-poor stellar population has been preferentially stripped, which would have implications for the mass--metallicity relation.

This paper is organized as follows. In Section \ref{sec:data}, we describe the \sfive data. Methods and quantitative results are presented in Sections \ref{sec:methods} and \ref{sec:results}, respectively. Section \ref{sec:discuss} is dedicated to discussing formation channels for Cra2 in the context of our findings. Finally, conclusions and a brief summary are provided in Section \ref{sec:conclusions}.

\begin{figure*}[pt!]
\centering
\includegraphics[width=1.75\columnwidth]{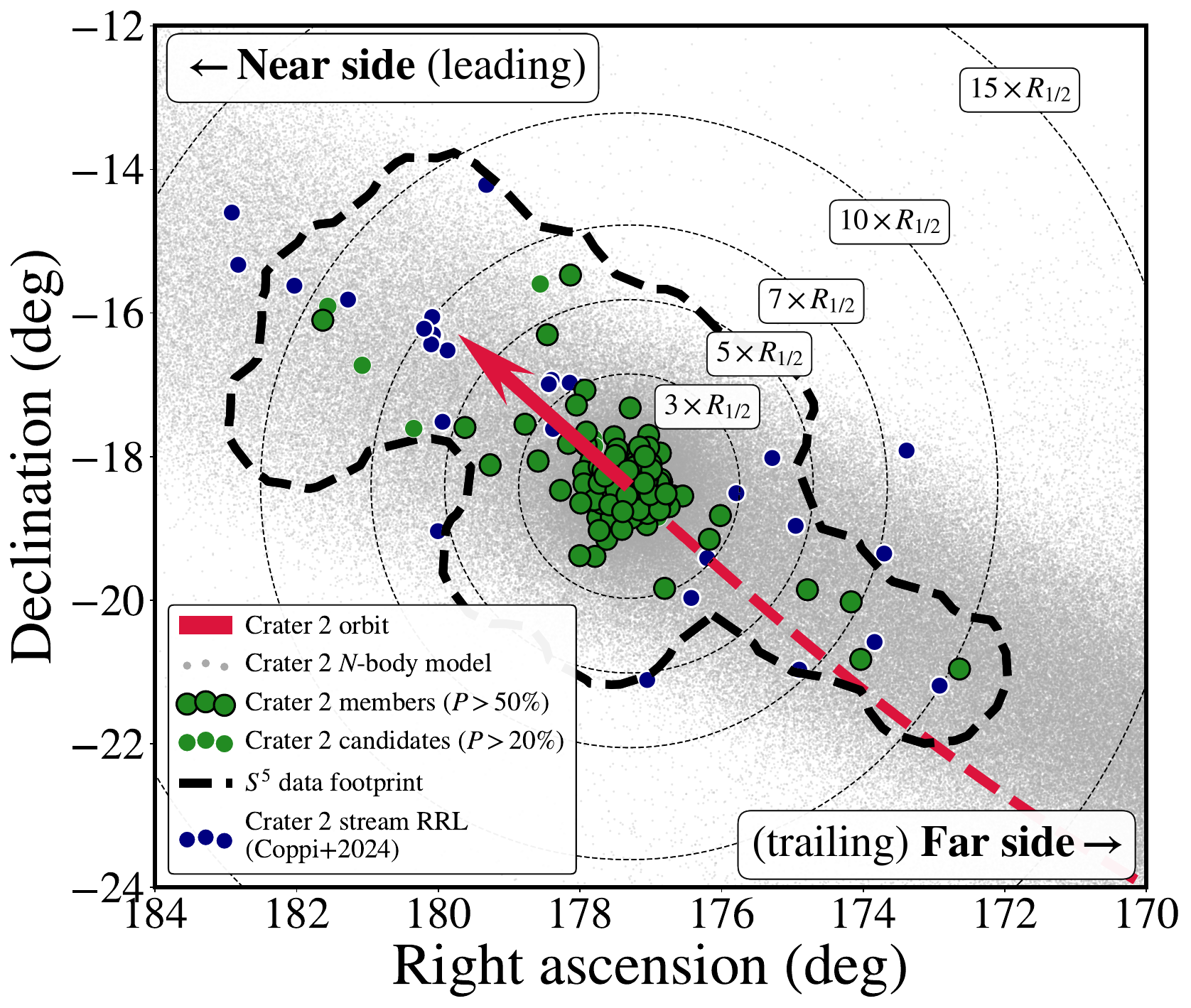}
\caption{On-sky stellar spatial distribution around the Crater~2 (Cra2) galaxy. Green circles with black edges are \sfive Cra2 members with a probability $P> 50\%$ from our mixture model (Section \ref{subsec:gmm}), including remnant body and stream. Other Cra2 candidates ($20\% < P \leq 50\%$) are shown as the smaller green dots. Variable RR Lyrae stars in the Cra2 stream are displayed as the dark blue circles with white edges \citep{coppi2024cra2}. The directions of both ``far side'' ($\alpha \lesssim 177.5^\circ$) and ``near side'' ($\alpha \gtrsim 177.5^\circ$) of the stream are annotated. 
The 2dF$+$AAOmega fields observed as part of \sfive are presented as the dashed black contour. The thin dotted lines mark 3, 5, 7, 10, and 15 $\rh$ around Cra2, where $\rh = 31.2$\,arcmin with a null ellipticity \citep{Torrealba2016cra2}. The Cra2 {\tt core-base} $N$-body model star particles (Section \ref{subsec:nbody}) are the gray dots in the background. Finally, the red arrow and dashed line exhibit the Cra2 proper motion vector in Galactic standard of rest frame and its past orbit used for the $N$-body simulations, respectively.
\label{fig:radec}}
\end{figure*}

\section{Data} \label{sec:data}



This paper utilizes data from the forthcoming $S^5$ second data release (DR2; T. S. Li et al., in preparation)\footnote{This corresponds to the internal release iDR3.7.}. The main science goal for $S^5$ is to characterize the kinematics and chemistry of stellar streams. For this, $S^5$ performs dedicated follow-up observations of known stellar stream fields with the dual-arm AAOmega spectrograph \citep{AAOmega} coupled with the Two-degree Field (2dF) fibre positioner \citep{2dF} at the 3.9\,m Anglo-Australian Telescope. $S^5$ initiated its program mid-2018 by targeting stellar streams identified in the Dark Energy Survey \citep[][and references therein]{Shipp2018}. Since then, $S^5$ has expanded to include several other stellar streams found in the Southern sky \citep[e.g.,][]{Li2022S5streams} as well as surviving, but tidally disrupting, dwarf galaxies and globular clusters.

In total, $S^5$ has observed 14 fields of $\sim$3\,deg$^2$ each around the center of Cra2 and along the predicted track for its stripped debris using the 2dF$+$AAOmega combination (Figure \ref{fig:radec}). Indeed, this debris track does coincide with the Cra2 stream detected by \citet{coppi2024cra2} with RRL stars. The central Cra2 field has been observed twice during February 2020 and the collected \sfive data have already been published in \citet{ji2021_cra2_ant2}. Two additional pointings with 2dF$+$AAOmega were taken on February 2021 positioned immediately around the central field along the stream track. The remaining 11 fields were observed during February and March 2023. Exposure times ranged from 2400 to 8400\,s depending on observing and/or weather conditions. The footprint geometry is neither uniform nor symmetric as 2dF failures have prevented us from filling out all the planned fields.

\begin{figure*}[pt!]
\centering
\includegraphics[width=2.1\columnwidth]{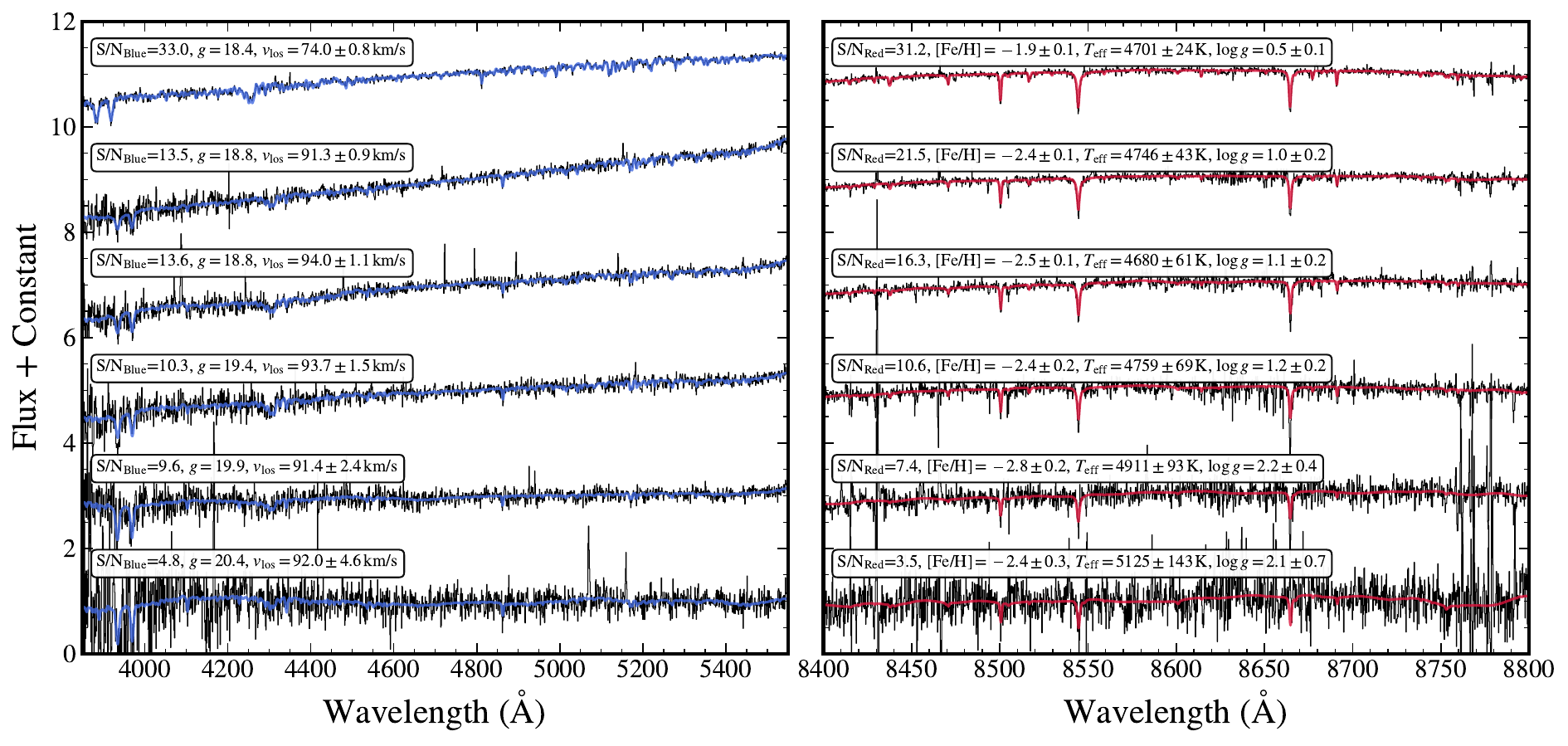}
\caption{Example 1D spectra for random Cra2 member stars (black). Left and right panels show blue- and red-arm AAOmega spectra, respectively. Best-fit model templates are overlaid as blue and red lines. On top of each exemplar, we list derived parameters with the \texttt{RVSpecFit} code from the \sfive~DR2 catalog, including blue- and red-arm signal-to-noise ratios ($S/N_{\rm Blue}$ and $S/N_{\rm Red}$, respectively), heliocentric line-of-sight velocity (\vhel), iron-to-hydrogen ratio metallicity (\feh), effective temperature (\teff), and logarithm of surface gravity acceleration (\logg). We also display DELVE~DR2 DECam $g$-band magnitude for each star (see text). From bottom to top, we order the spectra according to increasing $S/N_{\rm Red}$, which is also considered in the running text. The vertical axis shows the measured flux plus a constant offset to facilitate visualization.}
\label{fig:specplot}
\end{figure*}

Regarding target selection, we refer the interested reader to \citet{Li2019s5} for the general $S^5$ strategy. For Cra2 specifically, astrometric and photometric criteria were adopted. Candidates were vetted to reside within a 0.15\,mag color window around a dereddened \citep{Schlegel1998} empirical isochrone constructed based on known members of MW satellite galaxies \citep{Pace&Li2019} and shifted to the Cra2 distance modulus \citep[$20.35$, heliocentric distance of 117.5\,kpc;][]{Torrealba2016cra2} in a \textit{Gaia} \citep{GaiaMission} $G - G_{RP}$ vs. $G$ color--magnitude diagram \citep[CMD;][]{GaiaDR2Photometry, GaiaEDR3Photometry}. Targets were selected to also have proper motions consistent with the mean value for Cra2 \citep{Fritz2018dwarf_orbits}. Observations taken during 2020 (central fields) used astrometry and photometry from \textit{Gaia} DR2 \citep{gaiadr2}, while \textit{Gaia} (early)DR3 \citep{GaiaEDR3Summary, GaiaDR32023} was adopted for all others. \textit{Gaia}'s parallaxes were used to remove likely nearby MW contaminants by keeping those targets where $\texttt{parallax\_over\_error} < 3$\footnote{This quantity in the \textit{Gaia} astrometric catalog is simply the parallax measurement value divided by its standard error.}. Whenever some of the 392 2dF science fibers were still unused, priority was given to sources with proper motions consistent with Cra2, but outside the $G - G_{RP}$ color window. Within the 14 fields targeted by $S^5$, a total of 3838 sources were observed out of which 2757 were flagged as stars with good fits by our pipeline \citep[$\texttt{good\_star} = \texttt{TRUE}$ in the $S^5$ catalog;][]{Li2019s5}.

Details on the global data processing for \sfive can be found in other collaboration papers \citep{Li2019s5, Li2021AtlasAliqaUma, Li2022S5streams, Shipp2021massLMC, koposov2023orphan}. Briefly, we model blue and red arms of AAOmega spectra simultaneously with \texttt{RVSpecFit}\footnote{\url{https://github.com/segasai/rvspecfit}.} template fitting code \citep[][see also \citealt{koposov2011bootes1}]{rvspecfit2019} using a grid of stellar atmospheres and synthetic spectra from the PHOENIX library \citep{husser2013PHOENIX} interpolated using a neural network emulator, which provides smooth interpolation with continuous first and second derivatives, hence avoiding stellar parameters from accumulating at the discontinuity points of derivatives at the grid nodes \citep[for additional details, see][]{koposov2026desiDR1}. We also assume a photometric prior on effective temperature with both Dark Energy Camera (DECam; $g-r$ and $r-z$; \citealt{DECam2015}) and \textit{Gaia} ($G - G_{RP}$) colors as in \citet{Li2019s5}. The DECam photometry adopted is from the The DECam Local Volume Exploration Survey (DELVE) DR2 \citep{DELVE2021dr1, DELVEdr2}. This approach benefits from the wider wavelength coverage in the blue arm (3800--5800\,\AA), despite lower spectral resolution ($\mathcal{R} \sim 1300$), for better stellar parameter estimates while maintaining good line-of-sight velocity (\vhel) precision thanks to the higher resolution ($\mathcal{R} \sim 10{,}000$) around the calcium triplet (8400--8800\,\AA) in the red arm.

With the \texttt{RVSpecFit} processed data at hand, we perform several quality cuts, in addition to $\texttt{good\_star} = \texttt{TRUE}$, to ensure the reliability of \sfive DR2 parameters. First, we limit our sample to stars with signal-to-noise ratio $S/N > 3$ in the red arm. The typical \vhel uncertainty for RGB stars in \sfive DR2 is ${<}5\,\kms$ at this $S/N$. Figure \ref{fig:specplot} illustrates how the near-infrared calcium triple lines be accurately identified even in the noisier spectra, driving the high precision in \vhel. Despite these \vhel errors at the low-$S/N$ end of our data being being larger than the literature (and this work's) internal velocity dispersion of Cra2 (2.3--2.7\,$\kms$), this precision is still sufficient for separating true galaxy members from the MW's foreground. Moreover, Section \ref{subsec:gmm} below describes our Bayesian mixture modeling approach that naturally accounts for uncertainties in estimating stellar membership in Cra2 and deriving its bulk kinematic properties. Even so, we still remove a few stars with large \vhel errors of ${>}10\,{\rm km \ s^{-1}}$, but we confirm that our qualitative results about Cra2 remain even with a stricter $S/N > 10$ cut. Then, we exclude stars with very large \textit{Gaia} proper motions in either right ascension ($\alpha$) or declination ($\delta$) directions: $|\mu_\alpha \cos{\delta}| \ > 10\,{\rm mas \ yr^{-1}}$ or $|\mu_{\delta}| \ > 10\,{\rm mas \ yr^{-1}}$.

After these above-listed quality cuts, we convert \vhel values to the Galactic standard of rest (\vgsr) using MW fundamental parameters from \texttt{Astropy} \citep[v5.10;][]{astropy, astropy2018}. The distance from the Sun to the Galactic center is 8.122\,kpc \citep{GRAVITY2018}, the Sun's velocity vector in the Galactocentric cylindrical frame is $(V_R, V_\phi, V_z)_\odot = (12.9, 245.6, 7.78)$\,km\,s$^{-1}$ \citep{Drimmel2018sunVel}, and the Sun's vertical displacement with respect to the Galactic plane is 20.8\,pc \citep{Bennett&Bovy2019vertical}. We then keep only those stars where $|\vgsr| \ < 500\,{\rm km \ s^{-1}}$. For stars with large iron-abundance metallicity errors (${>}0.5\,{\rm dex}$), we set the [Fe/H] uncertainty to 99\,dex, but do not discard them from the sample. This allows us to keep their phase-space information for fitting the Cra2 stream kinematics (Section \ref{subsec:gmm}), while effectively disregarding them for metallicity determination. These bad [Fe/H], and \vhel, measurements in the \sfive catalog are usually due to issues with sky subtraction, causing large systematic residuals in the spectra. In total, 1100 stars with good-quality velocities, i.e., passing all the described cleaning criteria, are considered for Cra2 membership (Appendix \ref{appendix:data}).

\newpage
\section{Methods} \label{sec:methods}

\subsection{$N$-body models of Crater~2 and its stellar stream} \label{subsec:nbody}

For comparison to our new $S^5$ DR2 data, as well as the \citet{coppi2024cra2} RRL sample, we compute both cored and cuspy models of Cra2 in the combined potential of the MW and LMC. We run these models with a restricted $N$-body method, which is based on the technique described in \cite{Vasiliev2021tango}. In this approach, we initialize an $N$-body system with tracer particles and fit these with a low order multipole expansion. We then evolve the tracer particles in this potential until the time for the next potential update. At this moment in time, the multipole expansion is recomputed and the tracer particles are then evolved in the new potential. This method has been implemented in \texttt{AGAMA} \citep{agama}\footnote{\url{https://github.com/GalacticDynamics-Oxford/Agama/blob/master/py/tutorial_streams.ipynb}.} and it provides a good match to full $N$-body simulations at a fraction of the computational cost (A. Senkevich et al., in preparation).

\subsubsection{Basic settings}

Here, we describe the basic settings that are common to all simulations. For the MW, we use the axisymmetric \texttt{MWPotential2014} from \cite{Bovy2015}. Following \cite{Vasiliev2021tango}, we model the LMC as a truncated NFW profile \citep{NFW1996halos, nfw1997} with a mass of $1.5\times10^{11}\,\msun$ \citep[][]{Erkal2019massLMC} and a scale radius of $r_s = 10.84\,{\rm kpc}$. This choice of mass and $r_s$ gives a rotation curve at 8.7\,kpc consistent with the results from \citet{vanderMarel+2014}. We take the truncation radius to be $10 \times r_s$, equivalent to 108.4\,kpc. We include dynamical friction on the LMC by the MW as in \citet{Vasiliev2021tango} and also account for the Galaxy's reflex motion in response to the LMC infall.

For the present-day phase-space coordinates of Cra2, we take heliocentric proper motions from \citet{pace2022dwarfs}, heliocentric distance from the distance modulus of \citet{Torrealba2016cra2}, and systemic \vhel from \citet{ji2021_cra2_ant2}. We note that these choices for both proper motions and \vhel are 1$\sigma$ consistent with the values derived in this work (Section \ref{subsec:grads}). For the LMC, we assume proper motions measured from multi-epoch \textit{Hubble Space Telescope} observations \citep{Kallivayalil2013lmc}. For the LMC's distance, we utilize the result from the eclipsing-binary measurement by \citet[][49.97\,kpc]{Pietrzynski2013_LMCdist}. The LMC's systemic \vhel is from \citet[][262\,km\,s$^{-1}$]{VanDerMarel2002massLMC}. We then rewind Cra2 in the combined potential of the MW and LMC for 7\,Gyr, at which point it is close to apocenter. We then evolve the system to the present day using the restricted $N$-body method described above.

\subsubsection{Cuspy NFW models}

For the cuspy Cra2 models, we treat the dark matter halo with an NFW profile \citep{NFW1996halos, nfw1997}. We compute a base realization with total mass $10^{8.6}\,\msun$ and fiducial concentration set by the mass--concentration relation of \citet{Dutton+2014} assuming a Hubble constant of $H_0 = 67.4$\,km\,s$^{-1}$\,Mpc$^{-1}$ \citep[][]{PlanckCollab2020}. We also use an exponential truncation with a truncation radius of two virial radii to eliminate distant dark matter particles. For the stellar component, we model it as a Plummer sphere \citep{Plummer1911} with 
$r_s = 1.066\,{\rm kpc}$ and $\mstar = 3.16\times10^5\,\msun$ \citep[][and see Equation \ref{eq:mstar}]{ji2021_cra2_ant2}. To eliminate stars at unrealistically large radii, we include an exponential truncation with a cutoff radius of $10\times r_s$, i.e., 10.66\,kpc.
We model both stellar and dark matter components with $10^6$ particles each and generate the initial conditions in \texttt{AGAMA}. The final velocity dispersion of the dwarf remnant in this model is 5.92\,$\kms$ and we refer to it as `{\tt cusp-base}'.

In order to find a closer match to the present-day central velocity dispersion of Cra2 without completely destroying the simulated dwarf, we run a grid of alternative simulations with halo masses between $10^6{-}10^9\,\msun$ (in steps of 0.2\,dex), and with concentrations between 0.1--10 times the expected value from the \citet[][]{Dutton+2014} relation in steps of 0.333\,dex. We find that a cuspy NFW model with mass $10^{7.6} M_\odot$ and fiducial concentration exhibits the lowest present-day dispersion of 3.76\,$\kms$;  we refer to this model as `{\tt cusp-lowmass}'. Note that this lowest dispersion is still larger than the observed value of $\sim$2.3--2.7\,$\kms$ for Cra2 \citep[][including our own findings in Sections \ref{subsec:gmm} and \ref{subsec:grads}, see Table \ref{tab:galaxy_props}]{Caldwell2017cra2, ji2021_cra2_ant2}. Additional effort to explore a larger parameter space using the restricted $N$-body technique will be required to satisfactorily reproduce Cra2's cold internal kinematics.

\begin{figure}[pt!]
\centering
\includegraphics[width=1.0\columnwidth]{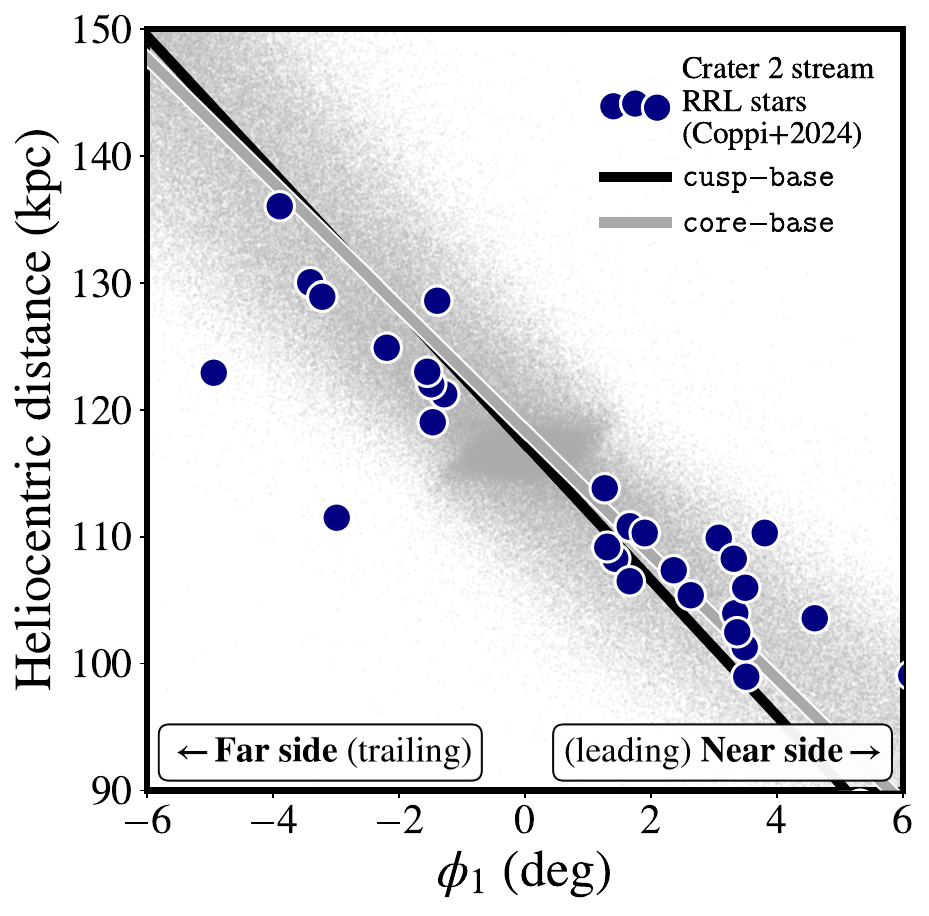}
\caption{Heliocentric distances as a function of stream longitude $\phi_1$ for the Crater~2 (Cra2) system. Dark blue symbols with white edges are RR Lyrae stars associated with the Cra2 stream \citep{coppi2024cra2}. The simulated Cra2 star particles from the {\tt core-base} model (Section \ref{subsec:nbody}) are shown as the small gray dots. The gray and black lines represent the $-4.9\,{\rm kpc}\,{\rm deg}^{-1}$ and $-5.4\,{\rm kpc}\,{\rm deg}^{-1}$ heliocentric distance gradients predicted by the {\tt core-base} and {\tt cusp-base} simulations, respectively.
\label{fig:dist_grad}}
\end{figure}

We note that the initial mass of the above-described {\tt cusp-lowmass} model is much lower than expected from abundance matching \citep[e.g.,][]{behroozi2013, Behroozi2019_UNIVERSEmachine, Rodriguez-Puebla2017}, including dwarf-galaxy specific stellar-to-halo mass relations \citep{Garrison-Kimmel2017shmr, read2017shmr, Jethwa2018simulations, nadler2020, Manwadkar2022grumpy, danieli2023shmr, ZaritskyBehroozi2023smhm, wang2024sagaUM, KadoFong2025shmr}. Hence, we check the model realizations with same initial mass as the {\tt cusp-base} simulation ($10^{8.6}\,\msun$) and verify that an NFW halo with low concentration (0.464 times smaller than the fiducial value) produces a central velocity dispersion almost identical (3.77\,$\kms$) to the {\tt cusp-lowmass} case \citep[for an exploration in this direction with Cra2, see][]{amorisco2019cra2}. We refer to this higher-mass low-concentration cuspy model as `{\tt cusp-lowc}'. 


\subsubsection{Cored halo models}

For the dark matter halos of the cored models, we use the \textsc{core}NFW prescription from \citet{read2016cores}. In particular, we consider a maximally cored scenario where the core radius is $1.75\times r_s$ 
of our Plummer sphere. We run a parameter grid search with masses in the range $10^7{-}10^9\,\msun$ and concentrations between 0.1--10 times the value set by the mass--concentration relation in \cite{Dutton+2014}. We find that a cored model with a mass of $10^{8.6}\,\msun$ and a concentration 0.464 times smaller than the fiducial concentration gives the smallest present-day velocity dispersion of 
4.79\,$\kms$ in comparison to other cored simulations. This cored model is dubbed `{\tt core-lowc}'. 

We also consider a model where the core radius is half the maximally cored value. For this latter cored case, we explore a similar parameter space and find that a mass of $10^{8.6}\,\msun$ and a fiducial concentration gives the smallest present-day velocity dispersion of 
4.21\,$\kms$ compared to the other simulations in this grid search. Given the identical initial mass and concentration to {\tt cusp-base}, we treat this as our `{\tt core-base}' model. As with the cuspy realizations, we generate initial conditions for these dwarfs with \texttt{AGAMA} and use $10^6$ particles for both stellar and dark matter components. We show a comparison between {\tt cusp-base} and {\tt core-base} models and the \citet{coppi2024cra2} RRL stars in Figure \ref{fig:dist_grad}. The full comparison between our $N$-body results and the \sfive DR2 data is in Section \ref{subsec:grads}.
\subsection{Stream coordinates} \label{subsec:coords}

Similarly to other stellar stream works where the progenitor is known, we define a convenient coordinate system for our analyses \citep[e.g.,][]{Majewski2003, Li2018tucanaIII}. We define the stream longitude $\phi_1$ and latitude $\phi_2$ in a way that $(\phi_1, \phi_2) = (0.0, 0.0)$ is located at the on-sky center of Cra2 ($\alpha = 177.310^\circ$ and $\delta = -18.413^\circ$; \citealt{Torrealba2016cra2}). The Cra2 stream is well aligned with $\phi_2 = 0$. We adopt a position angle of 60$^\circ$ to derive the $3 \times 3$ rotation ($R$) matrix that converts from celestial ($\alpha, \delta$) to stream coordinates ($\phi_1, \phi_2$):
\begin{equation}
\begin{bmatrix}
\cos{(\phi_1)} \cos{(\phi_2)} \\ 
\sin{(\phi_1)} \cos{(\phi_2)} \\ 
\sin{(\phi_2)}
\end{bmatrix} = R \times 
\begin{bmatrix}
\cos{(\alpha)} \cos{(\delta)} \\ 
\sin{(\alpha)} \cos{(\delta)} \\ 
\sin{(\delta)}
\end{bmatrix}.
\label{eq1}
\end{equation}
The rotation matrix entries $R_{ij}$ are presented to a precision of 8 significant digits,
\begin{linenomath}
\begin{dmath}
R = \begin{bmatrix}
R_{00} & R_{01} & R_{02} \\ 
R_{10} & R_{11} & R_{12} \\ 
R_{20} & R_{21} & R_{22}
\end{bmatrix} = \begin{bmatrix}
-0.94775886 & \phantom{-}0.04452939 & -0.31586432 \\ 
-0.19840253 & -0.85765902 & \phantom{-}0.47440218 \\ 
-0.24977905 & \phantom{-}0.51228716 & \phantom{-}0.82168869
\end{bmatrix},
\end{dmath}
\end{linenomath}
and aligns well with the \citet{coppi2024cra2} RRL stars. See \citet{Koposov2010gd1} and \citet{Shipp2019streamsPMs} for a similar formalism applied to other stellar streams. 


\subsection{Stellar kinematics and membership} \label{subsec:gmm}

To identify Cra2 members across the entire \sfive footprint, we model the galaxy's remnant body and stream kinematics in a Bayesian mixture modeling framework. This probabilistic approach has already been extensively employed to disentangle members of stellar streams, as well as dwarf galaxies and globular clusters, from their surrounding back/foreground \citep{martin2008sats, walker2009membership, walker2016_tuc2_gru1, Li2018tucanaIII, Pace&Li2019, pace2022dwarfs}, including \sfive works \citep{Shipp2019streamsPMs, Li2022S5streams, Awad2024gcsS5}. 
We model the distributions of \textit{Gaia} proper motions, including correlations between $\mu_\alpha \cos{\delta}$ and $\mu_\delta$, and \texttt{RVSpecFit} \vhel and \feh within \sfive DR2 Cra2 fields with three components. One of these corresponds to the actual Cra2 system, including galaxy and stream. The other two components represent the foreground contamination from the MW's disk and halo. For an unbiased identification of Cra2 members in its outskirts and stream, we do not consider spatial priors. 

\begin{figure*}[pt!]
\centering
\includegraphics[width=2.04\columnwidth]{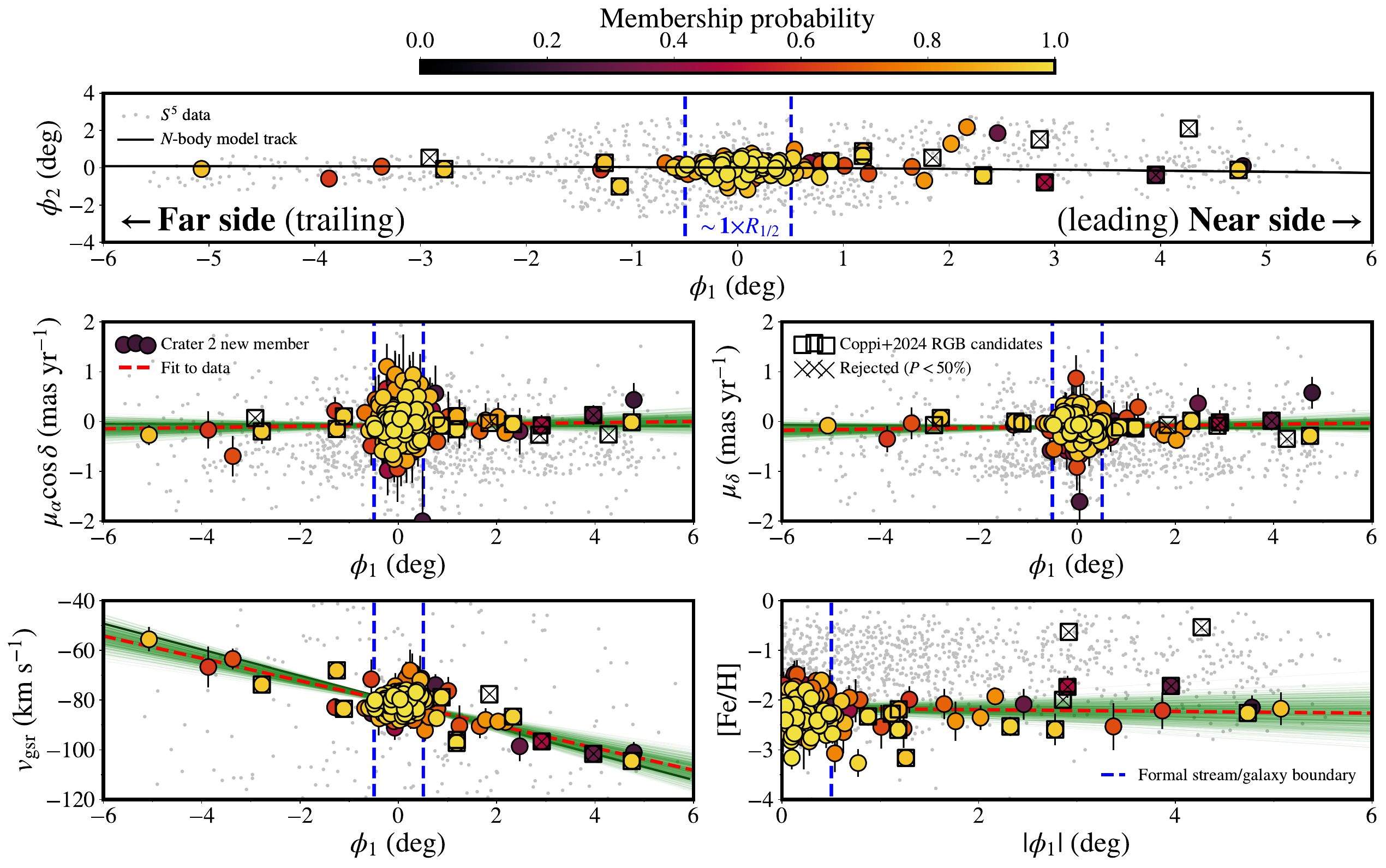}
\caption{Mixture modeling fit results. Top panel: Crater~2 (Cra2) stream coordinates $(\phi_1, \phi_2)$ distribution (Section \ref{subsec:coords}). The far (trailing arm) and near (leading) sides of the stream are annotated. Middle: linear proper motion gradients as a function of stream longitude $\phi_1$. Bottom left: linear Galactic standard of rest line-of-sight velocity ($v_{\rm gsr}$) gradient. Bottom right: radial metallicity (\feh) gradient. In all panels, Cra2 candidates ($P>20\%$) in the \sfive DR2 footprint are color-coded by their membership probabilities. Non-members that have also been observed by \sfive are plotted as gray dots. Red dashed lines represent the mean values from our best-fit mixture model while the thin green lines show 200 realizations of our fit sampled from our posterior chains. The black solid lines are linear fits to our {\tt core-base} simulated stream track (Section \ref{subsec:nbody}). Red giant-branch Cra2 stream candidates from \citet{coppi2024cra2} are plotted as black open squares. Those with available \sfive DR2 data, but rejected as true members, are displayed as black `X' symbols. In all panels, blue dashed lines delineate the formal boundary between galaxy and stream in our mixture modeling analysis (Section \ref{subsec:gmm}).
}
\label{fig:results}
\end{figure*}

\renewcommand{\arraystretch}{1.0}
\setlength{\tabcolsep}{0.3em}
\begin{table*}[ht!]
\centering
\caption{Crater~2 galaxy properties
}
\label{tab:galaxy_props}
\begin{tabular}{>{\normalsize}l >{\normalsize}l >{\normalsize}l >{\normalsize}l >{\normalsize}l}
\hline
Parameter & Value & Unit & Source & Description\\ %
\hline
\hline
$\alpha$ & 177.310 & deg & \citet{Torrealba2016cra2} &  Center right ascension \\ 
$\delta$ & $-$18.413 & deg & \citet{Torrealba2016cra2} & Center declination \\ 
$m-M$ & 20.35$\pm$0.02 & [mag] & \citet{Torrealba2016cra2} & Distance modulus \\ 
$\dhelio$ & 117.5$\pm$1.1 & kpc & \citet{Torrealba2016cra2} & Heliocentric distance \\ 
$M_V$ & $-8.2$ & [mag] & \citet{Torrealba2016cra2} & $V$-band absolute magnitude \\
\rh & $1054^{+93}_{-89}$ & pc & \citet{pace2025lvdb}\footnote{Local Volume Database with structural parameters from \citet{Torrealba2016cra2}.\label{refnote}} & Physical circularized half-light radius\\
\rh & $31.2\pm2.5$ & arcmin & \citet{Torrealba2016cra2} & Projected angular half-light radius\\
$\epsilon$ & ${<}0.1$\footnote{Upper limit at 95\% confidence level. We adopt $\epsilon =0.0$ in Figures \ref{fig:rhalf_mu} and \ref{fig:radec}.}  & & \citet{Torrealba2016cra2} & Ellipticity\\
$\mu_V$ & 30.5 & [mag] & \citet{pace2025lvdb}\footref{refnote} & Average surface brightness within \rh \\
$\mstar$ & $10^{5.5}$ & \msun & Equation \ref{eq:mstar} & Stellar mass\footnote{Assuming mass-to-light ratio $\eta = 2$ and the Sun's absolute $V$-band magnitude $M_{V,\odot} = +4.81$.} \\
\hline
$\mu_\alpha \cos{\delta}$ & $-$0.070$\pm$0.020 & mas\,yr$^{-1}$ & This work & Heliocentric proper motion ($\alpha$) \\
$\mu_\delta$ & $-$0.107$\pm$0.013 & mas\,yr$^{-1}$ & This work & Heliocentric proper motion ($\delta$) \\
\vhel & $+89.2\pm0.3$ & ${\rm km\,s^{-1}}$ & This work & Heliocentric line-of-sight velocity \\
\vgsr & $-81.2\pm0.3$ & ${\rm km\,s^{-1}}$ & This work & Galactic standard of rest line-of-sight velocity \\
$\sigma_{v,{\rm gal}}$ & $2.51^{+0.33}_{-0.30}$ & ${\rm km\,s^{-1}}$ & This work & Line-of-sight velocity dispersion within \rh \\
$\sigma_{v,{\rm str}}$ & $5.74^{+0.98}_{-0.83}$ & ${\rm km\,s^{-1}}$ & This work & Line-of-sight stream internal velocity dispersion \\
$\sigma_{v, {\rm str}}/\sigma_{v, {\rm gal}}$ & $2.30^{+0.41}_{-0.35}$ & & This work &  Velocity dispersion ratio between stream and galaxy\\
$\Delta \vgsr/\Delta \phi_1$ & $-4.5\pm0.6$ & ${\rm km\,s^{-1}}$\,deg$^{-1}$ & This work & \vgsr gradient as a function of stream longitude $\phi_1$ \\
$\langle \feh \rangle$ & $-2.16\pm0.04$ & [dex] & This work & Mean central metallicity \\
$\sigma_{\feh}$ & $0.28\pm0.03$ & [dex] & This work & Metallicity dispersion \\
\hline
$\Delta \dhelio/\Delta \phi_1$ & $-4.9$ & kpc\,deg$^{-1}$ & This work & $N$-body model \dhelio gradient with $\phi_1$\footnote{Average between all $N$-body models considered (Section \ref{subsec:nbody}).\label{refnote2}} \\
$r_{\rm apo}$ & 154.31& kpc & This work & $N$-body model orbital apocenter  \\
$r_{\rm peri}$& 22.26 & kpc & This work & $N$-body model orbital pericenter \\
$e$ & 0.75 & & $e=\dfrac{r_{\rm apo} - r_{\rm peri}}{r_{\rm apo} + r_{\rm peri}}$ & $N$-body model orbital eccentricity \\
\hline
\end{tabular}
\end{table*}

We model \vgsr, \feh, and proper motions as Gaussian distributions for both Cra2 (galaxy plus stream) and foreground components. The full likelihood functions are written elsewhere \citep{walker2009membership, walker2016_tuc2_gru1, Li2018tucanaIII, Pace&Li2019, Awad2024gcsS5}. Following the expectation from the stream $N$-body models (Section \ref{subsec:nbody}), we adopt a linear dependence of mean \vgsr with $\phi_1$ in the form
\begin{equation}
\vgsr (\phi_1) = \left(\dfrac{\Delta \vgsr}{\Delta \phi_1}\right) \phi_1 + \vgsr (\phi_1 = 0),
\end{equation}
where $\Delta \vgsr / \Delta \phi_1$ is the velocity gradient and $\vgsr (\phi_1 = 0)$ is Cra2's mean systemic \vgsr at $\phi_1 = 0$. We find a significant velocity gradient at the ${\approx}7\sigma$ level: 
\begin{equation}
\dfrac{\vgsr (\phi_1)}{\kms} = \dfrac{(-4.5 \pm 0.6)}{\rm deg} \phi_1 + (-81.2 \pm 0.3),
\end{equation}
where this systemic \vgsr is compatible (1$\sigma$) with previous results \citep{Caldwell2017cra2, fu2019cra2, ji2021_cra2_ant2}. Implications and comparisons with the $N$-body models are in Section \ref{subsec:grads}. 

For metallicity, we assume a linear radial gradient with $|\phi_1|$ in our mixture model. We have explored using a quadratic function instead, which might be appropriate for some dwarf galaxies \citep[see][]{taibi2022gradients}. However, we find no evidence for a significant [Fe/H] gradient in either case. 
The mean central metallicity ($\langle \feh \rangle = -2.16 \pm 0.04$) and dispersion ($\sigma_{\rm \feh} = 0.28\pm0.03$) we find are effectively identical to the values from \citet[][]{ji2021_cra2_ant2}. We also assume a linear dependence with $\phi_1$ for the proper motions, but found no significant evidence for variations. The mean systemic proper motions we estimate are, again, 1$\sigma$ consistent with other works \citep{Fritz2018dwarf_orbits, ji2021_cra2_ant2, battaglia2022propermotions, pace2022dwarfs}
; $\langle \mu_\alpha \cos \delta \rangle = -0.070\pm0.020$\,mas\,yr$^{-1}$ and $\langle \mu_\delta \rangle = -0.107 \pm 0.013$\,mas\,yr$^{-1}$. For the foreground components, we assume no spatial dependence for any of these parameters.

To account for Cra2's galaxy and stream kinematics separately, we assume that the fraction of members in the sample changes depending on $\phi_1$ \citep[similar to][]{Awad2024gcsS5}. Formally, our model incorporates a ``central fraction'' and a ``stream fraction'' terms which are defined inside and outside the boundary $|\phi_1| = 0.5^\circ$ \citep[$\sim$Cra2's \rh;][]{Torrealba2016cra2}. This inclusion is necessary since adopting a single fraction for the whole galaxy would artificially increase the membership probability ($P$) of stars in the stream while decreasing it for Cra2's central region. This stream--galaxy boundary also serves for us to estimate the velocity dispersion in the stream and remnant body separately. We present the full quantitative results in Section \ref{sec:results} (also Table \ref{tab:galaxy_props}).

We sample posterior distributions with a Markov chain Monte Carlo approach using the \texttt{emcee} package \citep{Foreman-Mackey2013emcee} with 200 walkers and 20,000 steps, including a burn-in stage of 10,000. We take $16^{\rm th}$ and $84^{\rm th}$ percentiles as uncertainties for all parameters, including the ones already listed in the previous paragraphs. The model fitting results are mostly presented in Figure \ref{fig:results} where Cra2 candidate stars at $P > 20\%$ are color-coded by their individual membership probability values. For the central velocity dispersion of Cra2, we find $\sigma_{v, {\rm gal}} = 2.51^{+0.33}_{-0.30}\,\kms$. For the stream, we obtained $\sigma_{v, {\rm str}} = 5.74^{+0.98}_{-0.83}$. As a sanity check, we also run fits with more stringent $S/N$ cuts. At $S/N > 5$, the quantitative results are identical to the ones in Table \ref{tab:galaxy_props}, at the $\sim$0.1\,$\kms$ level for $\sigma_{v, {\rm gal}}$. With $S/N > 10$, we estimate $\sigma_{v, {\rm gal}} = 3.1^{+0.9}_{-0.6}\,\kms$. Although this value is nominally larger than our fiducial estimate, we note that the uncertainty is much larger due the smaller number of stars going into the fit ($<$30\% of the original sample), so this is still consistent within 1$\sigma$. Perhaps more important, the ratio between $\sigma_{v, {\rm str}}$ and $\sigma_{v, {\rm gal}}$ remain $\approx$2.1. Hence, our qualitative conclusions and their implications to the formation of Cra2 and other diffuse dwarfs (Section \ref{sec:discuss}) remain unchanged.

Out of the 1100 stars with good \texttt{RVSpecFit} parameters in the \sfive DR2 catalog within the footprint that pass all quality criteria at $S/N > 3$ (Section \ref{sec:data}), we identify 143 Cra2 members ($P > 50\%$), of which 114 belong to the main body of the galaxy \citep[within its \rh;][]{Torrealba2016cra2, Vivas2020cra2} and 29 are associated with the stellar stream. From these numbers, we can naively estimate a lower limit for Cra2's \mstar loss of $21\pm7\%$ \citep[Wilson score interval at 95\% confidence level;][]{Wilson1927}. Cra2 star members in the \sfive footprint can be found out to ${\gtrsim}10{\times}$ its on-sky angular \rh ($>$20\,kpc physical separation), aligned in a stream-like morphology that matches the expected debris track (Figure \ref{fig:radec}). Note that we observe Cra2 stars across the entire \sfive coverage in the $\phi_1$ direction and, therefore, are yet to reach the edge of the stellar stream. Our mixture modeling results, as well as other Cra2 galaxy properties, are listed in Table \ref{tab:galaxy_props}. 


\section{Results} \label{sec:results}

The most obvious takeaway from our analysis of \sfive DR2 data 
is the unequivocal detection of a stellar stream attached to the Cra2 dwarf galaxy remnant and, therefore, that this system is experiencing severe tidal disruption. Below, we list other results alongside pertinent discussions.

\subsection{Discovery of red giant-branch stars in the Crater~2 stream } \label{subsec:rgb}

\begin{figure}[pt!]
\centering
\includegraphics[width=1.0\columnwidth]{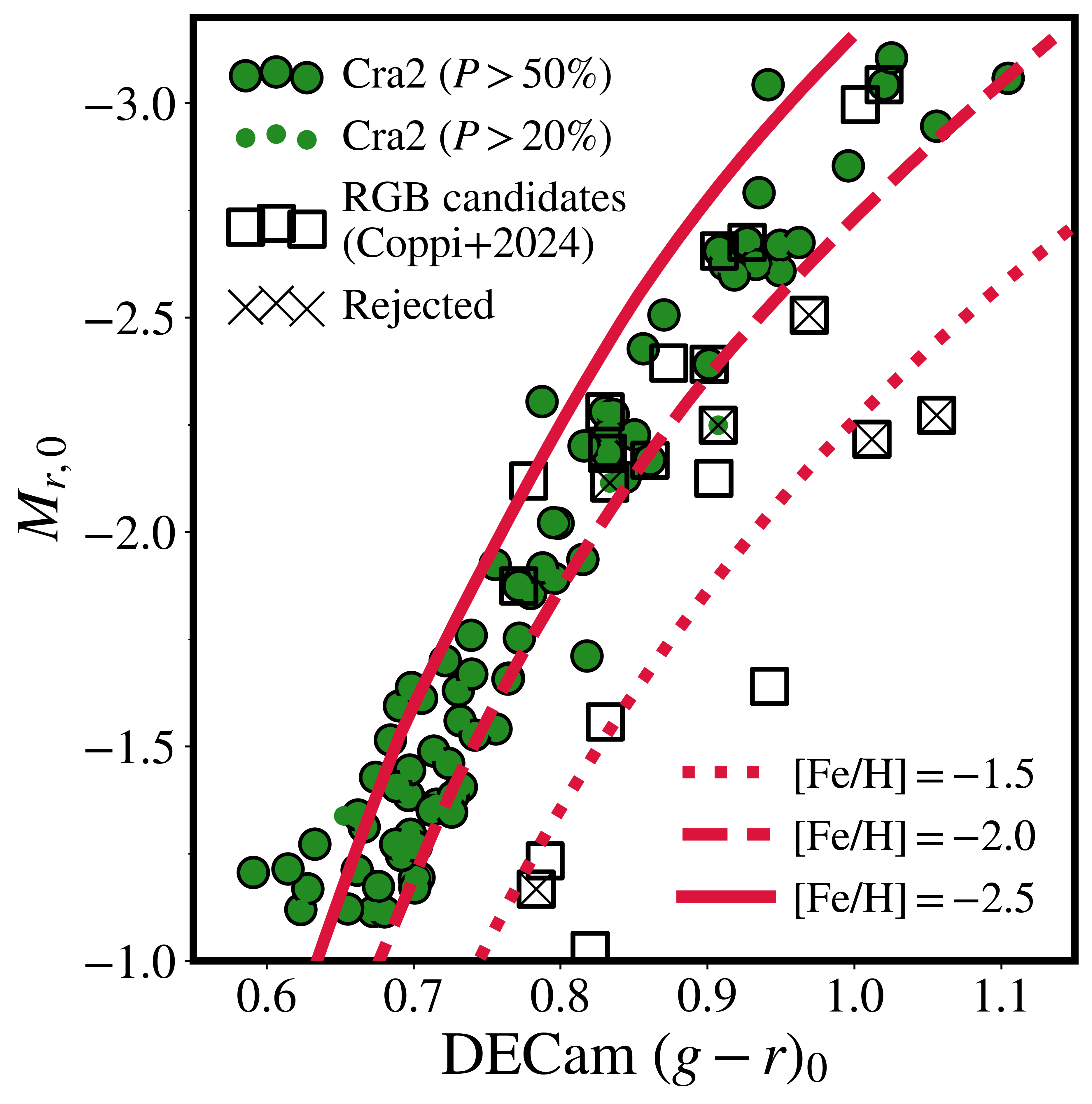}
\caption{Dereddened DECam $(g-r)_0$ vs. $M_{r,0}$ color--magnitude diagram covering the red giant branch (RGB) with photometry from DELVE DR2 \citep{DELVEdr2} for stars in the \sfive DR2 footprint with Crater~2 (Cra2) membership probability $P>50\%$ (green circles with black edges) and $20\%<P\leq50\%$ (green dots). The absolute $r$-band magnitude is computed assuming the expected distance gradient from our Cra2 {\tt core-base} model (Section \ref{subsec:nbody}). The \citet{coppi2024cra2} Cra2 RGB candidates are plotted as the open squares, with the rejected ones having an `X' symbol on top. Red lines are (12.5\,Gyr, $[\alpha/{\rm Fe}] = +0.4$) Dartmouth isochrones \citep{dotter2008} with $\feh = -2.5$ (solid), $-2.0$ (dashed), and $-1.5$ (dotted).
\label{fig:cmd}}
\end{figure}

In comparison to previous \sfive work \citep[][]{ji2021_cra2_ant2}, we recover almost all members within Cra2's central region. The few stars that were considered members by \citet{ji2021_cra2_ant2}, but are now attributed $P < 50\%$, are usually the most metal-rich ones in the sample at $\feh \gtrsim -1.5$. Out of our 143 members, 114 are located within $|\phi_1| \ < 0.5^\circ$ (${\sim}$Cra2's $\rh$). The remaining 29 stars are formally members of the stream in our mixture model. The farthest Cra2 stream star within the \sfive DR2 footprint 
is found 
at ${\sim}10{\times}$ Cra2's on-sky \rh and $\gtrsim$20\,kpc physical separation given the expected distance gradient
. As can be appreciated from Figure \ref{fig:cmd}, the majority of the new Cra2 RGB members are contained within very metal-poor ($\feh = -2.0$ and $-2.5$) Dartmouth isochrones \citep{dotter2008} of old stellar populations (12.5\,Gyr, $[\alpha/{\rm Fe}] = +0.4$) in a DECam $g-r$ vs. absolute $r$-band magnitude ($M_r$) CMD \citep[DELVE DR2 photometry;][]{DELVEdr2}. Absolute magnitudes for our sample are computed assuming the heliocentric distance predicted for Cra2 tidal tails from our {\tt core-base} stream model (Figure \ref{fig:dist_grad}). The narrow color range covered by Cra2's RGB is in line with the 
deep CMD analysis by \citet{walker2019cra2}.

A total of 24 potential RGB members were identified in the Cra2 stream by \citet{coppi2024cra2}. Out of these, 14 are located within the \sfive DR2 footprint and were observed as part of the program (square symbols in Figure \ref{fig:radec}). We confirm that 8 of these RGB stars are, indeed, associated with the Cra2 stream with high probability, all at $P \geq 88\%$ (Table \ref{tab:coppi}). All these new Cra2 stream stars exhibit very low metallicities ($-3.2 \lesssim \feh < -2.2$) and \vgsr values consistent with the predicted {\tt core-base} stream track (bottom left panel of Figure \ref{fig:results}). Some of the other candidates are rejected due to their discrepant \vgsr (CS4, WS2, and WS5; $P=0.0$ in Table \ref{tab:coppi}), which are even outside the plot boundaries in the $\phi_1$ vs. \vgsr panel in Figure \ref{fig:results}. Star CS12 is $\sim$20$\kms$ displaced from the fitted $\vgsr(\phi_1)$ track at the same time that it has a very high metallicity, though also quite uncertain, in the \sfive DR2 catalog ($\feh \approx +0.4\pm0.5$). The rest of the \citet{coppi2024cra2} candidates (CS13, and WS4) are attributed $30\% \leq P < 50\%$ likely due to their relatively high metallicities ($\feh \approx -1.7$).


Despite the above-described overall success of \citeauthor{coppi2024cra2}'s (\citeyear{coppi2024cra2}) method for finding Cra2 stream members, there was still some contamination. We verify that all the rejected candidates from these authors' sample turned out to be those redder, in DECam $g-r$ color, than most of the Cra2 members found in \sfive DR2. In the DECam $g-r$ vs. $M_r$ CMD shown in Figure \ref{fig:cmd}, the rejected Cra2 RGB candidates are all redder than expected from the $\feh = -2$ Dartmouth isochrone, some being redder than the $\feh = -1.5$ track. The \sfive spectroscopically vetted members are almost entirely confined within, or bluer than, the $\feh = -2.0$ and $-2.5$ isochrones. Hence, narrowing the color window around Cra2's RGB sequence and adding information regarding the predicted distance gradient might be advantageous when searching for additional stream members even farther away from this galaxy's center
.

\renewcommand{\arraystretch}{1.0}
\setlength{\tabcolsep}{0.5em}
\begin{table*}[ht!]
\centering
\caption{Crater~2 stream red giant-branch candidates from \citet{coppi2024cra2} with \sfive DR2 observations
}
\label{tab:coppi}
\begin{tabular}{ccccccccccc}
\hline
\textit{Gaia} DR3 \texttt{source\_id} & Name & $\alpha$ & $\delta$ & $g^\dagger$ & $r^\dagger$ & \vhel & \vgsr & \feh & $P$ \\
& (\citealt{coppi2024cra2}) & (deg) & (deg) & & & (km\,s$^{-1}$) & (km\,s$^{-1}$) & &  \\
\hline
3546585982161428736 & CS4$\phantom{0}$ & 174.3575 & $-$19.3930 & 19.3 & 18.2 & $\phantom{00}$2.6$\pm$1.3 & $-$175.7 & $-$0.64$\pm$0.10 & 0.00 \\
3542009329433731072 & CS5$\phantom{0}$ & 174.7955 & $-$19.8505 & 19.7 & 18.8 & 104.7$\pm$3.6 & $-$73.8$\phantom{0}$ & $-$2.59$\pm$0.32 & 0.94 \\
3543686119025619200 & CS6$\phantom{0}$ & 176.0250 & $-$18.8178 & 18.8 & 17.8 & 105.8$\pm$1.2 & $-$68.1$\phantom{0}$ & $-$3.16$\pm$0.11 & 0.93 \\
3542608597631389568 & CS7$\phantom{0}$ & 176.8130 & $-$19.8322 & 18.8 & 17.8 & $\phantom{0}$91.3$\pm$1.0 & $-$83.4$\phantom{0}$ & $-$2.27$\pm$0.10 & 0.96 \\
3544050091734226304 & CS8$\phantom{0}$ & 177.9139 & $-$17.6618 & 18.9 & 18.0 & $\phantom{0}$88.7$\pm$1.2 & $-$78.8.1$\phantom{0}$ & $-$2.33$\pm$0.18 & 0.98 \\
3567352213460408960 & CS9$\phantom{0}$ & 179.6337 & $-$17.5965 & 18.9 & 18.1 & $\phantom{0}$77.4$\pm$1.3 & $-$86.5$\phantom{0}$ & $-$2.52$\pm$0.19 & 0.97 \\
3568269210454897792 & CS10 & 178.0557 & $-$17.2831 & 18.4 & 17.3 & $\phantom{0}$69.2$\pm$0.9 & $-$97.1$\phantom{0}$ & $-$2.59$\pm$0.07 & 0.98 \\
3568288486268051712 & CS11 & 177.9339 & $-$17.0760 & 18.9 & 18.1 & $\phantom{0}$70.2$\pm$1.3 & $-$95.9$\phantom{0}$ & $-$2.19$\pm$0.19 & 0.88 \\
3568206877592789888 & CS12 & 178.7070 & $-$17.0423 & 19.2 & 18.1 & $\phantom{0}$86.9$\pm$1.2 & $-$77.6$\phantom{0}$ & $+$0.42$\pm$0.53 & 0.19 \\
3567764324162907520 & CS13 & 181.0822 & $-$16.7292 & 18.9 & 18.0 & $\phantom{0}$57.4$\pm$1.3 & $-$101.5 & $-$1.72$\pm$0.17 & 0.30 \\
3567951176715579008 & CS14 & 181.6388 & $-$16.1012 & 18.8 & 17.9 & $\phantom{0}$51.8$\pm$1.1 & $-$104.4 & $-$2.26$\pm$0.16 & 0.96 \\
3568907335219100032 & WS2$\phantom{0}$ & 179.0963 & $-$15.6769 & 19.9 & 19.0 & 266.8$\pm$2.0 & $+$106.4 & $-$1.99$\pm$0.19 & 0.00 \\
3567316960369373696 & WS4$\phantom{0}$ & 180.3542 & $-$17.6119 & 18.9 & 18.0 & $\phantom{0}$66.0$\pm$2.3 & $-$96.5$\phantom{0}$ & $-$1.73$\pm$0.23 & 0.46 \\
3570603950380122496 & WS5$\phantom{0}$ & 180.0465 & $-$14.4625 & 18.9 & 17.8 & 127.7$\pm$1.7 & $-$27.6$\phantom{0}$ & $-$0.54$\pm$0.15 & 0.00 \\
\hline
\end{tabular}
\begin{flushleft}
$^\dagger$DECam $g$ and $r$ photometry reported in DELVE DR2 \citep{DELVEdr2}.
\end{flushleft}
\end{table*}

\newpage
\subsection{Kinematics along the Crater~2 stellar stream \\ and insights on its dark matter content} \label{subsec:grads}

An important immediate byproduct of our kinematic analysis is our $\approx$7$\sigma$ detection and determination of a velocity gradient across the Cra2 system; $\Delta \vgsr / \Delta \phi_1 = (-4.5 \pm 0.6)\,\kms\,\deg^{-1}$. This quantity can be used as an additional constraint to future tailored simulations of Cra2 trying to reproduce its structural properties. A comparison between \sfive DR2 new Cra2 RGB members and the stream {\tt core-base} model track, alongside our linear fit to the $\vgsr$ gradient, is presented in the bottom left panel of Figure \ref{fig:results}. The output $\vgsr (\phi_1)$ values from our $N$-body simulations are in reasonable agreement with observations within the on-sky coverage of \sfive DR2, the range of model velocity gradients being within ${-}4.6$ and ${-}5.4\,\kms\,\deg^{-1}$.

In dark matter-only tidal disruption simulations, \citet{Errani2015tidalEvo} showed that the velocity-dispersion ratio between stream and central region in a dwarf galaxy ($\sigma_{v, {\rm str}}/\sigma_{v, {\rm gal}}$) is sensitive to its dark matter density profile. In general, disrupted cored dark matter halos will exhibit higher values of $\sigma_{v, {\rm str}}/\sigma_{v, {\rm gal}}$ than cuspy ones. The reason for that is because, to reproduce a fixed value of central velocity dispersion ($\sigma_{v, {\rm gal}}$), cored halos require a larger total mass than cuspy ones. As a result, the tidal radii, i.e., the sizes, of cored dark matter halos are larger than cuspy ones, leading to dynamically hotter and spatially wider tidal streams. Furthermore, dwarf galaxies embedded in cored dark matter halos cool down more efficiently during tidal stripping than their cuspy counterparts, leading to increasing $\sigma_{v, {\rm str}}/\sigma_{v, {\rm gal}}$ over time. Note that hydrodynamical simulations predict that most dwarf satellites around MW-like hosts should be tidally disrupting, but their associated stellar streams would have surface brightnesses that are too faint to be detected in current photometric data \citep{shipp2023fireStreams, shipp2025aurigaStreams, riley2025aurigaMZR}. Therefore, future kinematic studies of yet-undiscovered stellar streams around seemingly unscathed MW satellites might be crucial for inferring their true dark matter mass profiles.



According to the \citet{Errani2015tidalEvo} calculations for a dwarf galaxy on an orbit of similar eccentricity to Cra2 \citep[$e \sim 0.7$;][]{fu2019cra2, ji2021_cra2_ant2, pace2022dwarfs}, an NFW cuspy halo would produce $\sigma_{v, {\rm str}}/\sigma_{v, {\rm gal}} \leq 1$ while the cored case gives $\sigma_{v, {\rm str}}/\sigma_{v, {\rm gal}} \gtrsim 3$. The ratio we estimate from the \sfive DR2 Cra2 data is $\sigma_{v, {\rm str}}/\sigma_{v, {\rm gal}} = 2.30^{+0.41}_{-0.35}$. At face value, this result disfavors an NFW cuspy halo, thus supporting the hypothesis of a dark matter core in Cra2. We also fit \sigv values for both Cra2 stream and remnant separately with only high-probability members ($P>50\%$) and found an almost identical ratio of $\sigma_{v, {\rm str}}/\sigma_{v, {\rm gal}} = 2.41^{+0.81}_{-0.64}$, confirming the robustness of this measurement.

Unfortunately, the direct comparison of the \citet{Errani2015tidalEvo} models to the \sfive data has several limitations. For instance, these authors' simulations do not account for the infalling LMC system. 
Additionally, these authors' model satellites reach equilibrium quickly after a pericentric passage. Although Cra2 is already $\sim$2\,Gyr past pericenter in our orbital configuration, an impulsive perturbation, such as a close encounter with the MW, may cause a dwarf galaxy to undergo a long-lived damped-oscillation phase with short-period velocity dispersion variations that could affect the $\sigma_{v, {\rm str}}/\sigma_{v, {\rm gal}}$ measurement \citep{errani2025_impulsive_oscillations}.It is also unclear how the galaxy's self-gravity might affect the stream kinematics close to the progenitor. Moreover, \citet{Errani2015tidalEvo} focuses on the velocity dispersion perpendicular to the stream plane, whereas \sfive velocities are line-of-sight measurements, which, despite accounting for the on-sky \vhel gradient in our mixture modeling, might still contain projection effects of both the in-plane and perpendicular motions. Perhaps the most critical caveat is that the \citet{Errani2015tidalEvo} predictions for stream kinematics were taken a lot farther away from their simulated dwarf galaxies' centers (from 20 to 100\,kpc) than the \sfive DR2 coverage for Cra2.

\begin{figure*}[pt!]
\centering
\includegraphics[width=2.1\columnwidth]{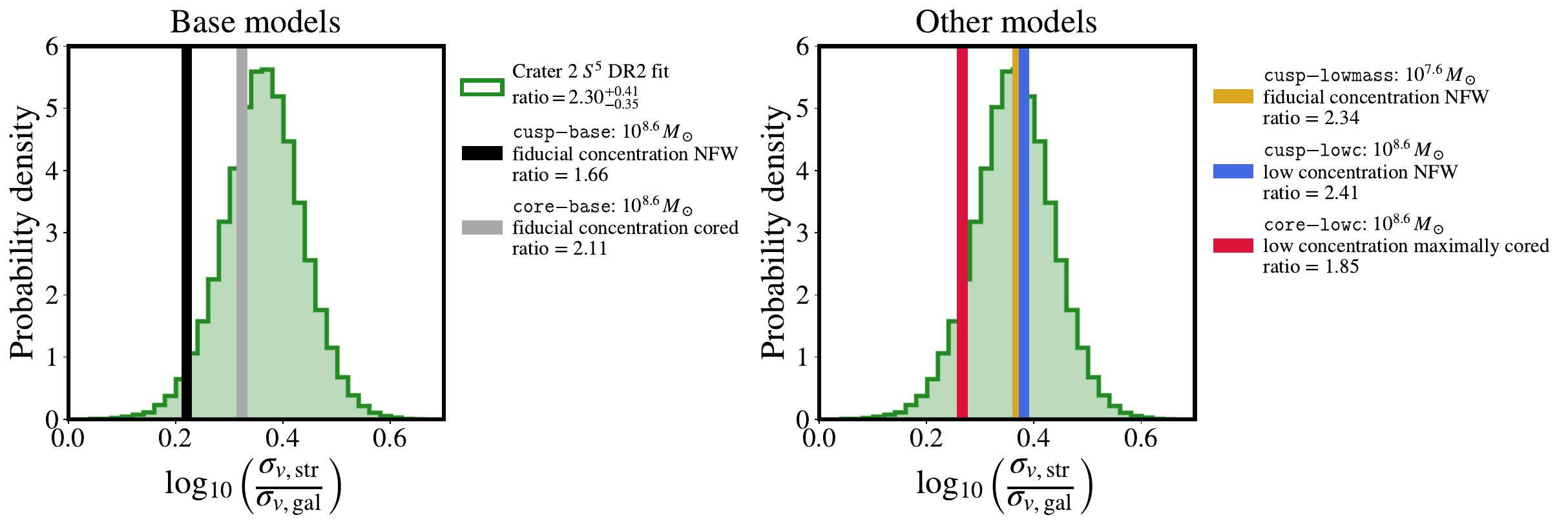 }
\caption{Distribution of velocity dispersion ratio between Cra2 stellar stream and remnant galaxy ($\sigma_{v, {\rm str}}/\sigma_{v, {\rm gal}}$) in logarithmic scale sampled from our mixture modeling results applied to the \sfive DR2 data (green histograms). Vertical lines represent the results for the velocity dispersion ratio of the different $N$-body models described in Section \ref{subsec:nbody} using a similar method as for the \sfive data. Left panel shows the comparison with the base models ($10^{8.6}\,\msun$ halo mass, fiducial concentration), namely {\tt cusp-base} (NFW cuspy, black line) and {\tt core-base} (cored, gray). Right panel shows the alternative models (see text for details): {\tt cusp-lowmass} ($10^{7.6}\,\msun$, fiducial concentration NFW, yellow), {\tt cusp-lowc} ($10^{8.6}\,\msun$, low concentration NFW, blue), and {\tt core-lowc} ($10^{8.6}\,\msun$, low concentration maximally cored, red). 
\label{fig:vel_disp}}
\end{figure*}

Here, we test if our own tailored cuspy and cored $N$-body models (Section \ref{subsec:nbody}) can reproduce the kinematics of Cra2's stream and galaxy. Although they do not exactly reproduce the present-day phase-space properties of the Cra2 system (e.g., central velocity dispersion), these models should provide qualitative insights on their behavior. In this vein, our $N$-body experiments show that halo concentration has an effect on the system's final kinematics degenerate with initial mass. The {\tt cusp-lowmass} \citep[$10^{7.6}\,\msun$, fiducial concentration;][]{Dutton+2014} and {\tt cusp-lowc} ($10^{8.6}\,\msun$, lower concentration) models reached identical central velocity dispersions despite the latter being 10$\times$ more massive, echoing results by \citet{amorisco2019cra2}. 

In our suite of $N$-body simulations, the limiting factor on how dynamically cold the model dwarfs can be is the complete disruption of satellites if halo mass and/or concentration are too low. For the cored halo case, however, not only halo mass and concentration are relevant, but the core radius also plays a role. The remnant velocity dispersion of the {\tt core-lowc} model \citep[maximally cored scenario;][]{read2016cores} is larger than in the {\tt core-base} model (half the core radius value) despite the lower concentration of the former.

To quantitatively compare Cra2's observed kinematics to the simulation results, we fit line-of-sight velocities as Gaussian distributions for both galaxy and stream in all $N$-body models with \texttt{emcee} \citep{Foreman-Mackey2013emcee} accounting for spatial gradients (results in Figure \ref{fig:vel_disp}). This method is analogous to that applied to the \sfive DR2 data (Section \ref{subsec:gmm}), but without the need for MW foreground components. The {\tt cusp-lowmass}, {\tt cusp-lowc}, and {\tt core-base} models can all reproduce (1$\sigma$) the velocity dispersion ratio between Cra2 stream and remnant body measured from \sfive DR2 data (see text above). The {\tt cusp-base} and {\tt core-lowc} models show lower ratio values, but still within $\lesssim$2.0$\sigma$ and 1.5$\sigma$, respectively. We recall that the {\tt cusp-lowmass} model has an unrealistically low initial mass compared to abundance matching expectations in the dwarf-galaxy regime \citep[e.g.,][]{Garrison-Kimmel2017shmr, read2017shmr, Jethwa2018simulations, nadler2020, Manwadkar2022grumpy, danieli2023shmr, ZaritskyBehroozi2023smhm, wang2024sagaUM, KadoFong2025shmr}. Hence, similar to previous works, we judge that an NFW halo progenitor with fiducial concentration is unlikely for Cra2. A broader exploration with $N$-body simulations covering additional parameters, such as MW potential, Cra2's orbit and stellar mass, and core radius size, will be required to obtain robust conclusions. Nevertheless, we find that both cuspy profiles with low concentration and cored halos with small core radius, compared to the maximally cored case \citep{read2016cores}, are promising avenues to explain Cra2's unique properties.

\subsection{Crater~2 is metal-poor compared to the Local Group stellar mass--metallicity relation} \label{subsec:mzr}

\begin{figure*}[pt!]
\centering
\includegraphics[width=2.0\columnwidth]{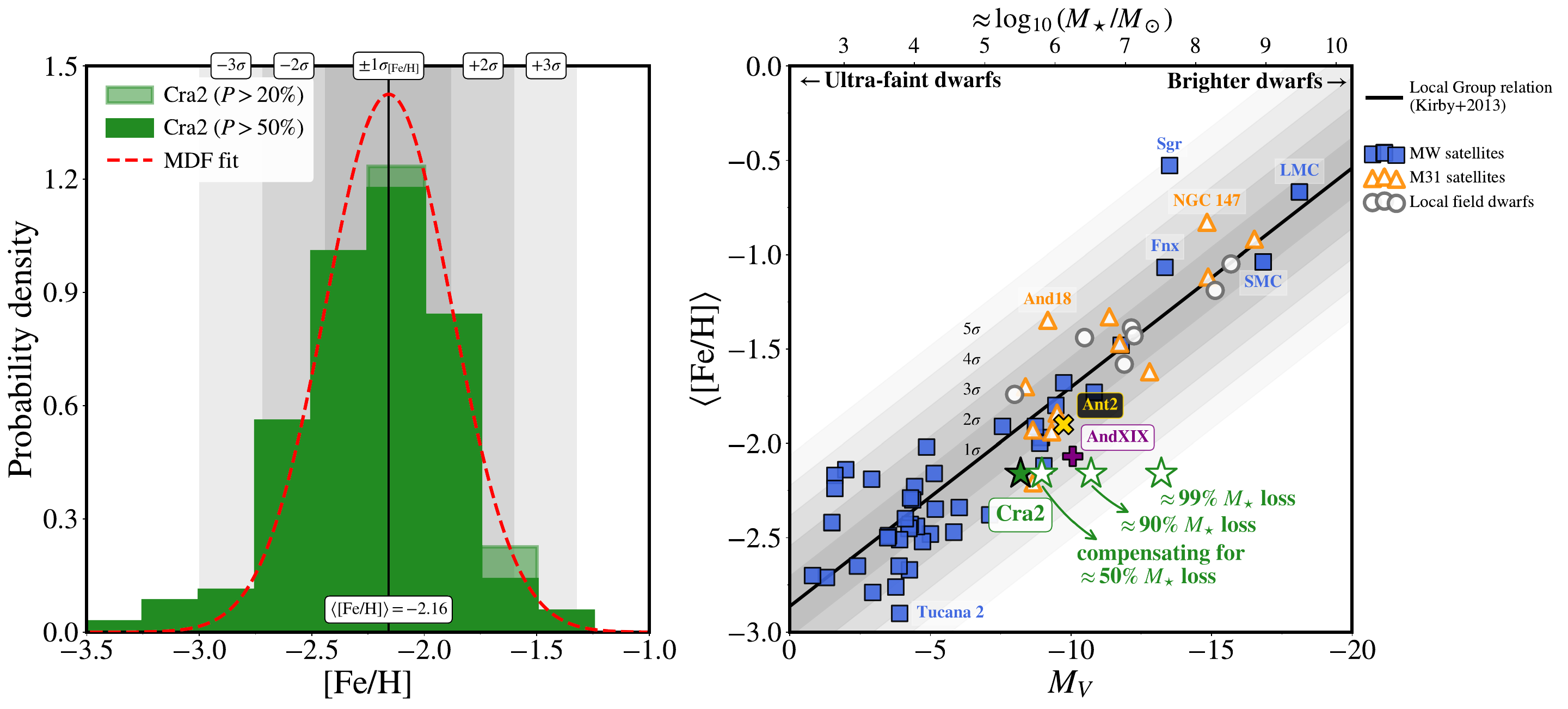}
\caption{Left: Crater~2 (Cra2) metallicity distribution function (MDF) for membership probability $P>50\%$ stars (green histogram). The transparent green histogram shows the same, but for $P>20\%$. The binning is set to 0.25\,dex, which is compatible with the \sfive DR2 typical \feh uncertainties for red giant-branch stars. Our Gaussian fit to the Cra2 MDF is represented by the red dashed line. Gray vertical bands illustrate $\pm$1, 2, and 3 times the $\sigma_{\feh} = 0.28$ ($\pm$0.03) dispersion around the $\langle \feh \rangle = -2.16$ ($\pm$0.04) black line. Right: stellar mass--metallicity relation (MZR) for Local Group dwarf galaxies. Visual scheme is the same as Figure \ref{fig:rhalf_mu}. Diffuse dwarfs Cra2 (green `$\star$' symbol), Antlia~2 (yellow `X'), and Andromeda~19 (purple `$+$') are highlighted with metallicities from this work, \citet{ji2021_cra2_ant2}, and \citet{Collins2020andXIX}, respectively. Milky Way satellites, M31 satellites, and Local Group field dwarf galaxies are plotted as blue squares, orange triangles, and gray circles, respectively. Chemical information is from \citet{Kirby2013} and \citet{simon2019}, with the exception of the Large and Small Magellanic Clouds (LMC and SMC) for which we take values from \citet[][see text]{Limberg2022gse}. Absolute $V$-band magnitude values are from the \citet{pace2025lvdb} compilation. The corresponding stellar masses ($M_\star$) are computed as in Equation \ref{eq:mstar}. We include Cra2's location in the horizontal $M_V$ axis assuming different $M_\star$ loss scenarios (white `$\star$' symbols with green edges). The MZR from \citet{Kirby2013} is the black line with $\pm$1, 2, 3, 4, and 5$\sigma$ scatter as the gray bands.
\label{fig:mzr_mdf}}
\end{figure*}

Another byproduct of our mixture modeling is the determination of the mean metallicity of Cra2 to be $\langle \feh \rangle = -2.16 \pm 0.04$ (full metallicity distribution in left panel of Figure \ref{fig:mzr_mdf}), adding to the census of Local Group dwarf galaxies with chemical information. With the current data, we find no evidence for \feh variations as function of $\phi_1$ across the Cra2 system (bottom right panel of Figure \ref{fig:results}). Metallicity gradients are commonly found in Local Group dSph galaxies \citep[e.g.,][]{taibi2022gradients}, including distinguishable chemo-dynamical populations \citep{pace2020ursaminor, ArroyoPolonio2024scl}. However, the latter usually requires metallicities for $>$1000 stars in a given dSph, while we only identify 143 Cra2 members (at $P>50\%$) in the \sfive DR2 footprint. Given the large on-sky extent of Cra2, a plausible strategy to test for the presence of spatial \feh changes might be to use wide-field photometric metallicities (as in \citealt[][]{Barbosa2025scl} for Sculptor dSph and \citealt{AlmeidaFernandes2026sgr} for Sagittarius dSph).

We now compare Cra2's mean metallicity derived from \sfive DR2 data to the demographics of Local Group dwarfs by placing this galaxy in the MZR of \citet[][also \citealt{simon2019}]{Kirby2013}. For consistency, most metallicity information for the sample of dwarf galaxies in the right panel of Figure \ref{fig:mzr_mdf} is either from \citet{Kirby2013} or \citet{simon2019}. The only exceptions are the LMC and SMC for which average \feh values were obtained based on the members from \citet{Nidever2020} with updated data from the Apache Point Observatory Galactic Evolution Experiment \citep[APOGEE;][DR17]{apogee2017, APOGEEdr17} by \citet{Limberg2022gse}. All the $M_V$ values are from the \citet{pace2025lvdb} compilation. We homogeneously compute corresponding \mstar for the displayed $M_V$ interval:
\begin{equation} \label{eq:mstar}
\dfrac{M_\star}{M_\odot} = \eta 10^{0.4(M_{V,\odot} - M_V)},
\end{equation}
where $\eta = 2$ is the assumed mass-to-light ratio and $M_{V, \odot} = +4.81$ \citep{willmer2018solar_absM}. Average metallicities for other diffuse dwarfs Ant2 and And19 are from \citet[][$\langle \feh \rangle = -1.90$]{ji2021_cra2_ant2} and \citet[][$-2.07$]{Collins2020andXIX}, respectively.

Arguably, the most intriguing observational fact about the collective of diffuse dwarfs in the Local Group is that they are all metal-poor compared to the expectation from the Local Group MZR for their luminosity \citep{Kirby2013, simon2019}. Ant2, Cra2, and And19 are located $\sim$1.0, 1.5, and 2.0$\sigma$, respectively, below the MZR assuming the root mean square scatter of 0.16\,dex around $\langle \feh \rangle$ derived by \citet{Kirby2013}. We note that, despite both slope and intercept of the linear relation from \citet{simon2019} being identical to \citeauthor{Kirby2013}'s (\citeyear{Kirby2013}), the scatter found by the former is inflated due to the presence of ultra-faint dwarf (UFD) galaxies ($\mstar \lesssim 10^5\,\msun$) in their fit, so we consider the latter. Although the diffuse dwarfs do not individually represent major deviations from the MZR, it is certainly unusual that all of them have lower-than-expected metallicities, hinting at a shared formation/evolution channel.

The location of diffuse dwarf galaxies relative to the Local Group MZR becomes more noteworthy since we are now positive that at least Cra2 is experiencing severe tidal disruption. When a galaxy loses stellar content, it should end up appearing metal-\textit{rich} compared to the MZR \citep{riley2025aurigaMZR}; such is the case for Sagittarius dSph \citep[e.g.,][see ``Sgr'' annotation in the right panel of Figure \ref{fig:mzr_mdf}]{Hayes2020}. Also, the Sagittarius dSph has experienced preferential stripping of its low-\feh population \citep[][]{Limberg2023sgr, cunningham2024sgr}, which further contributes to its displacement from the MZR\footnote{With APOGEE DR17, the mean metallicities of Sagittarius dSph remnant galaxy and stream are $\langle \feh \rangle = -0.58$ and $-1.07$, respectively \citep{Hayes2020, Limberg2022gse}} \citep{riley2025aurigaMZR}. Contrary to Sagittarius dSph, Cra2 is metal-\textit{poor} relative to the MZR and accounting for \mstar loss exacerbates such discrepancy. Previous works have argued that, to reproduce the large \rh and small $\sigma_v$ of Cra2 in an NFW dark matter halo, it would require $>$90\% loss of stellar content through tides \citep[e.g.,][]{Fattahi2018satellite_sims,applebaum2021DC_justLeague_sims}. Compensating for 90\% (99\%) \mstar loss results in an $M_V$ difference of 2.5 (5.0)\,mag, which would bring the tension with the MZR to the 3$\sigma$ (5$\sigma$) level (white `$\star$' symbols with green edges in Figure \ref{fig:mzr_mdf}).

In the spirit of the above discussion, the fact that And19 is metal-poor relative to the Local Group MZR was actually used as an argument against the tidal disruption mechanism to explain this galaxy's large size by \citet[][but see results by \citealt{Cullinane2024_AndXIX}]{collins2022andXIX}. Confirming And19's orbit around M31 with direct proper-motion measurements will be crucial to confirm (or falsify) that this galaxy is not experiencing mass loss through tides
. However, given the counter-example of Cra2, we can now assert that low metallicity and tidal disruption are not mutually exclusive. 

There is evidence that Ant2 is also being tidally disrupted \citep{ji2021_cra2_ant2}, but its deviation from the MZR is less extreme than Cra2's and And19's (see Figure \ref{fig:mzr_mdf}). In fact, \citet{riley2025aurigaMZR} has recently argued based on hydrodynamical galaxy-formation simulations that up to $\sim$80\% \mstar loss is still plausible for a displacement from the MZR similar to that observed in Cra2 and Ant2, but more than that, as required in the case of a cuspy NFW halo \citep{Fattahi2018satellite_sims, applebaum2021DC_justLeague_sims}, is difficult to explain. Indeed, one way to alleviate this tension between the diffuse dwarfs and the MZR might come from the evidence for either a cored density profile or a low-concentration NFW halo; for a fixed enlargement factor, these models require
less total mass loss than cuspy ones with fiducial concentration \citep{Errani2015tidalEvo, amorisco2019cra2, fu2019cra2}.




\newpage
\section{Discussion} \label{sec:discuss}

\citet{collins2022andXIX} discussed possible formation mechanisms for the unusual properties of And19, which are all in common with Cra2: ($i$) the colossal size for its \mstar, hence extremely low surface brightness \citep{Torrealba2016cra2}, ($ii$) minuscule central velocity dispersion \citep{Caldwell2017cra2, fu2019cra2}, ($iii$) lower-than-expected metallicity in comparison to the Local Group MZR \citep[][also Section \ref{subsec:mzr}]{ji2021_cra2_ant2}, and ($iv$) short star formation history (SFH) with quenching $\gtrsim$10\,Gyr ago \citep{walker2019cra2}. We build on these arguments, but incorporate our own findings, to scrutinize possible scenarios for the origin of Cra2's remarkable properties and, by extension, Ant2 and And19 as well. We also consider that ($v$) Cra2 is unmistakably tidally disrupting (Figures \ref{fig:radec} and \ref{fig:results}), including our detection of a stellar stream around it \citep[also][]{coppi2024cra2}, and ($vi$) the evidence that a cuspy NFW halo with fiducial concentration is unlikely to reproduce the system's kinematics with reasonable initial mass \citep[Figure \ref{fig:vel_disp} and Section \ref{subsec:grads}, as well as previous works;][]{sanders2018cra2, fu2019cra2, Borukhovetskaya2022cra2, errani2022}. Below, we consider if proposed formation pathways can fulfill this set of criteria.

\subsection{Dry mergers}
According to \citet{collins2022andXIX}, merger activity could be a mechanism to form And19 and, for the sake of the present discussion, the other diffuse dwarfs Cra2 and Ant2. These authors argued that dry mergers could naturally produce both the unusual structural properties of And19 and its low metallicity. In their hypothesis, the building blocks of And19, hence Cra2 and Ant2, would have been UFD galaxies since these are very metal-poor systems ($\langle \feh \rangle < -2$, see MZR in Figure \ref{fig:mzr_mdf}), commensurate with the diffuse dwarfs. Interestingly, since UFDs are essentially reionization fossils quenched at high redshift \citep[e.g.,][]{weisz2014sfhs}, this proposition would automatically explain the old stellar populations ($\gtrsim$10\,Gyr) of Cra2, And19, and Ant2. Simulation work by \citet{orkney2021coresMergersUFDs} has also claimed that late minor mergers could produce constant-density cores in dwarf galaxies of similar mass to Cra2 and And19. We note, however, that mergers with even fainter systems would not explain the formation of Cra2's stellar stream.



Indeed, galaxies in the mass range of Cra2 ($\mstar \sim 3\times10^5\,\msun$) and Ant2 ($\mstar \sim 10^6\,\msun$; $M_V = -9.7$ in Equation \ref{eq:mstar}) are expected to have experienced $\mathcal{O}(10)$ accretion events with even smaller systems \citep[e.g.,][]{griffen2018}. However, the same is true for all other classical MW satellites of similar luminosity such as Draco and Sculptor dSph galaxies and, therefore, it is difficult to invoke dry mergers as the sole responsible for the extreme properties of diffuse dwarfs. In any case, we mention this possibility since it can actually be tested with detailed chemical abundances for stars in the MW diffuse satellites. \citet{collins2022andXIX} suggested this possibility in the context of And19, but this galaxy is too distant \citep[$\sim$800\,kpc;][]{savino2022and}, hence its stars are too faint, for this kind of study with current instrumentation. Nevertheless, the most luminous RGB stars in Cra2 and Ant2 ($\sim$120--140\,kpc) are bright enough (DECam $g \sim 18.5$) that they are amenable to high-resolution spectroscopy ($\mathcal{R} \geq 20{,}000$) from the ground with 6--8\,m-class telescopes. More specifically, a lack of neutron-capture elements is a defining feature of UFDs \citep{ji2019ufds, simon2019} and can be used as a diagnostic.

\subsection{Star formation and supernovae feedback}
We have shown that a plausible explanation to the large velocity dispersion ratio between Cra2's stellar stream and remnant body is a cored halo with small core radius. Bursty star formation is usually accepted as the culprit for the core-cusp transformation in relatively bright ($\mstar \gtrsim 10^7\,\msun$) dwarf galaxies (see reviews by \citealt{Bullock2017review} and \citealt{Sales2022rev}). In this process, supernovae outflows transport the central gas outward and the associated gravitational potential fluctuations transfer energy to the collisionless dark matter particles, which end up in dynamically-hotter (higher-energy) orbits at larger radii \citep{navarro1996cores, ReadGilmore2005cores, pontzen2012cusp-core, brooks2014cores, El-Zant2016corecusp, read2016cores}. 

The effectiveness of the above-described feedback-induced core creation mechanism depends on the amount of energy to be injected into the galaxy's interstellar medium (the number of supernovae $\propto$ \mstar) and the total dark matter mass that needs to be moved around ($\propto$ halo mass \mvir), i.e., fundamentally the stellar-to-halo mass ratio \citep[$\mstar/\mvir$,][]{penarrubia2012corecusp, DiCintio2014cores}. Hydrodynamical simulations converge on a threshold where feedback becomes inefficient at producing cores of $\log (\mstar/\mvir) \lesssim -3.5$ \citep{tollet2016coresNIHAO, fitts2017coresFIRE, Benitez-Llambay2019coresEAGLE}, which translates to $\mstar \lesssim 10^6\,\msun$ using the empirical abundance matching relation from \citet{Rodriguez-Puebla2017}, which is about the \mstar of Sculptor dSph as well as Ant2.

Following the above-cited premise of abundance matching \citep[from][]{Rodriguez-Puebla2017}, the expectation for a galaxy of Cra2's \mstar is $\log (\mstar/\mvir) \approx -4.0$ prior to tidal disruption, which is well within the regime where core creation is not expected, at least via bursty star formation and supernovae feedback. We find almost identical values for \mstar/\mvir using the dwarf-galaxy specific present-day (redshift $z<0.05$) stellar-to-halo mass relation from \citet{KadoFong2025shmr} and the MW-tailored galaxy--halo connection prescription from \citet{nadler2020}. It is unclear, however, if we should consider the stellar-to-halo mass relation at redshift $z \sim 0$ or at the moment of Cra2's quenching $\sim$10.5\,Gyr ago ($z=2$ in \citealt{PlanckCollab2020} cosmology). If we adopt the latter, $\log (\mstar/\mvir) \approx -3.9$ for Cra2. In the most optimistic case, where we, again, compensate for 50\% \mstar loss in Cra2, we can bring it to $\log (\mstar/\mvir) \approx -3.7$ at redshift $z=2$. 

In addition to high $\mstar/\mvir$, recent works have pointed out that multiple cycles of star formation, i.e., a prolonged SFH of $\gtrsim$8\,Gyr \citep{Read2019cores, muni2025} are also required to form a dark matter core. Due to the early quenching of And19, \citet{collins2022andXIX} argued against a cored halo in that galaxy, a reasoning that also applies to Cra2. In Cra's CMD, two well-defined episodes of star formation can be identified, but at $\sim$10.5\,Gyr and $\sim$12.5\,Gyr in the past \citep{walker2019cra2}. However, we note that Sculptor dSph also experienced a short SFH \citep{Kirby2011alphas, deBoer2012sculptor, vincenzo2016sfhSculptor, betinelli2019sculptorSFH, delosReyes2022} and there are claims that this galaxy has a cored halo \citep[][but see \citealt{StrigariFrenkWhite2010, StrigariFrenkWhite2017} and \citealt{vitral2025scl}]{battaglia2008scl, walker2011dsph, amorisco2012scl, agnello2012sculptor}. Hence, confirming the true shape the dark matter density profile in Sculptor dSph with larger kinematic data sets will also be important to better constrain the core-cusp transformation process.

To summarize, the situation is the following. Given the low \mstar, hence low $\mstar/\mvir$ from abundance matching, and early quenching of Cra2, star formation and supernovae feedback should be insufficient to significantly reduce the central dark matter density and convert a cuspy NFW halo into a cored one. This line of reasoning might disfavor the cored halo hypothesis for Cra2, thus supporting a low-concentration NFW profile. However, the plausibility of this scenario largely depends on the scatter of the mass--concentration relation in the low-mass regime, which is difficult to constrain given the limiting resolution in cosmological simulations. Another justification for a low-concentration halo could be warm dark matter, 
which naturally predicts low-mass halos to be less concentrated than CDM ones \citep[e.g.,][and references therein]{maccio2013cdm+wdm, ludlow2016mass-concentration}. Having said that, if one is to invoke an interpretation outside collisionless CDM, solutions that would naturally produce a low-density central core have also been proposed for Cra2, including, e.g., self-interacting dark matter \citep[][]{Zhang2024sidmCra2}.



\subsection{High gas fraction and ram-pressure stripping}

Although stellar/supernovae feedback is typically assumed to be responsible for gas removal in dwarf galaxies, hence gravitational potential variations and redistribution of dark matter particles, other baryonic processes could, in principle, be responsible for the destruction of central cusps \citep[see discussion by][]{LiZhaozhou2023cusp-core}. Based on observations of bright post-starburst dwarfs ($10^8 < \mstar/\msun \lesssim 10^9$, similar to LMC and SMC) containing tails of stripped material in both galaxy clusters and groups, \citet{Grishin2021ram-pressure} proposed that gas removal via ram-pressure stripping could be responsible for the formation of UDGs. In this scenario, ram pressure would induce a star-formation burst followed by quick quenching. Also, the associated fast change in the gravitational potential due to gas loss would cause considerable expansion.

The collective of diffuse dwarfs in the Local Group, for which Cra2 is now the best-studied exemplar, could represent a more extreme version of the ``future-UDGs'' analyzed by \citet{Grishin2021ram-pressure}. In the case of an initially high gas fraction, in comparison to the average atomic hydrogen (\textsc{Hi}) mass--\mstar scaling relation \citep{huang2012alfafa, guo2021h1-mstar}, a large portion of these galaxies' total mass would be susceptible to ram-pressure stripping. During infall into the Local Group, the interaction with the MW's or M31's hot corona \citep[i.e., their circumgalactic medium,][]{tumlinson2017cgmReview} could have stripped the gas content from these galaxies \citep[e.g.,][]{NicholsBland-Hawthorn2011dwarfs}; these diffuse dwarfs have no evidence for current presence of gas or star formation in them. So, the associated gravitational potential variations, combined with tides, would have transformed the structural properties of these galaxies into their current diffuse status \citep[see also][]{wang2024hydroSimsDwarfs}. 

We finish by saying that the above-described hypothesis of high gas fraction combined with ram-pressure stripping is consistent with the low metallicity of Cra2 and other diffuse dwarfs (Ant2, And19). Since the number of supernovae occurring in a galaxy is proportional to its \mstar, the excess of gas would provide extra material to dilute metals into, diminishing the overall metallicity of these systems. This is similar to how star-formation rate (i.e., gas availability; \citealt{Kennicutt1998sf}) is anti-correlated with metallicity in the fundamental MZR \citep{mannucci2010fundMZR, Maiolino2019cosmicChemEvo}
. Future high-resolution hydrodynamical simulations should be able to test our conjectures.


\section{Conclusions} \label{sec:conclusions}

Below, we list our main results from the analysis of \sfive DR2 data in the fields around Crater~2 (Cra2). Then, we close the paper with a high-level summary paragraph.


\begin{itemize}
    \item Within the \sfive program \citep{Li2019s5}, we performed spectroscopic observations across 14 fields of $\sim$3\,deg$^2$ each around the diffuse MW dSph satellite Cra2 using 2dF$+$AAOmega. Out of these, 1100 genuine stars with good-quality data were considered for Cra2 membership (Appendix \ref{appendix:data}).

    \item We computed $N$-body models for the Cra2 system including a realistic past orbit in the combined potential of the MW and LMC. The simulations can broadly reproduce the properties of the Cra2 stellar stream, including a distance gradient of $-4.9\,{\rm kpc}\,{\rm deg}^{-1}$ that matches the distribution of variable RR Lyrae stars discovered by \citet{coppi2024cra2} and a velocity gradient consistent with the measurement from \sfive DR2 data.

    \item We fitted kinematics and chemistry of the Cra2 galaxy and stream with a Bayesian mixture modeling approach. Our results for the galaxy's main body are compatible with previous measurements, including a systemic line-of-sight velocity in the Galactic standard of rest frame of $\vgsr = -81.2 \pm 0.3\,\kms$ (heliocentric $\vhel = +89.2\pm0.3\,\kms$), central line-of-sight velocity dispersion of $\sigma_{v,{\rm gal}} = 2.51^{+0.33}_{-0.30}\,\kms$, and mean central metallicity of $\langle \feh \rangle = -2.16\pm 0.04$. Previous works could not measure a velocity gradient, whereas we found a $\approx$7$\sigma$ detection of $-4.5\pm0.6\,\kms\,{\rm deg}^{-1}$ along the Cra2 system.

    \item We identified 143 Cra2 red giant-branch (RGB) members at probability $P > 50\%$ within the \sfive DR2 footprint, including stars as far as 10 $\times$ Cra2's angular \rh, which translates to $\gtrsim$20\,kpc physical separation after accounting for the distance gradient. We formally classified 114 of these RGB stars as central members of the Cra2 galaxy and 29 as part of the stream. We then estimated a lower limit for Cra2's stellar mass loss of $21\pm7\%$. 

    \item Inspired by \citet{Errani2015tidalEvo}, we measured the velocity dispersion ratio between the Cra2 stream and central remnant to be $\sigma_{v, {\rm str}}/\sigma_{v, {\rm gal}} = 2.30^{+0.41}_{-0.35}$. We found that this observation is difficult to conciliate with an NFW cuspy halo with fiducial concentration and reasonable initial mass from abundance matching. Instead, either a cored halo with small core radius (see Section \ref{subsec:nbody}) or a low-concentration NFW model can reproduce this ratio. Additional $N$-body simulations covering a broader range of parameter space will be required to exactly match all phase-space properties of Cra2.


    \item Contrary to the expectation for a galaxy that has experienced tidal stripping \citep{riley2025aurigaMZR}, we found that Cra2 is $\approx$1.5$\sigma$ metal-poor compared to the \citet{Kirby2013} Local Group stellar mass--metallicity relation (MZR). Compensating for 50/90/99\% stellar mass loss brings this tension with the MZR to the 2/3/5$\sigma$ level. We concluded that tidal disruption and low metallicity are not mutually exclusive in diffuse dwarfs such as Cra2 (also Antlia~2 and Andromeda~19).
\end{itemize}

This paper demonstrates the power of spectroscopic observations at large radii around dwarf galaxies in the Local Group. We have shown that the diffuse Milky Way satellite Cra2 is undeniably experiencing tidal disruption and has a lower-than-expected metallicity in comparison to the MZR. Moreover, we have provided evidence that an NFW cuspy halo with fiducial concentration struggles to reproduce the combined kinematics of the Cra2 galaxy and stellar stream with reasonable halo mass from abundance matching. In the future, tidal tails might be key to explore the dark matter content in MW satellites from yet-to-be discovered low-surface brightness features around them. Also, future theoretical work must be able to fully replicate the properties of Cra2 and its siblings Antlia~2 and Andromeda~19. Otherwise, this emerging population of diffuse dwarf galaxies could represent a novel small-scale challenge to galaxy formation models and, perhaps, point to physics beyond the collisionless cold dark matter theory.










\software{\texttt{Anaconda} \citep{anaconda2020}, \texttt{CMasher} \citep{cmasher}, \texttt{corner} \citep{corner2016}, \texttt{gala} \citep{gala2017}, \texttt{jupyter} \citep{jupyter2016}, \texttt{matplotlib} \citep{matplotlib}, \texttt{NumPy} \citep{numpy}, \texttt{pandas} \citep{pandas2010}, \texttt{SciPy} \citep{scipy}, \texttt{scikit-learn} \citep{scikit-learn}, \texttt{TOPCAT} \citep{TOPCAT2005}.
}

\begin{acknowledgments}
This work is part of the ongoing \sfive program (\url{https://s5collab.github.io}).
\sfive includes data obtained with the Anglo-Australian Telescope in Australia. We acknowledge the traditional owners of the land on which the AAT stands, the Gamilaraay people, and pay our respects to elders past and present.

We thank the anonymous reviewer for comments and suggestions that contributed to this work. G.L. acknowledges funding from KICP/UChicago through a KICP Postdoctoral Fellowship. G.L. is indebted to all those involved with the multi-institutional \textit{MilkyWayBR} Group for the weekly discussions. G.L. also thanks Evan Kirby, Cristina Chiappini, and Marcel Pawlowski for useful comments during visits to University of Notre Dame/USA and to Leibniz-Institut f\"ur Astrophysik Potsdam/Germany as well as Chin Yi Tan for conversations over many coffee breaks and meetings at KICP. A.P.J. acknowledges support from the National Science Foundation under grant AST-2307599 and the Alfred P. Sloan Research Fellowship. T.S.L. acknowledges financial support from Natural Sciences and Engineering Research Council of Canada (NSERC) through grant RGPIN-2022-04794. D.E. thanks Eugene Vasiliev for helpful discussions about \texttt{AGAMA}. S.K. acknowledges support from the Science \& Technology Facilities Council (STFC) grant ST/Y001001/1. S.L.M. acknowledges support from ARC DP220102254 and the UNSW Scientia Fellowship program.

This work has made use of data from the European Space Agency (ESA) mission {\it Gaia} (\url{https://www.cosmos.esa.int/gaia}), processed by the {\it Gaia} Data Processing and Analysis Consortium (DPAC, \url{https://www.cosmos.esa.int/web/gaia/dpac/consortium}). Funding for the DPAC has been provided by national institutions, in particular the institutions participating in the {\it Gaia} Multilateral Agreement.

This work also made use of DELVE survey data. The DELVE project is partially supported by the NASA Fermi Guest Investigator Program Cycle 9 No. 91201. This work is partially supported by Fermilab LDRD project L2019-011. This material is based upon work supported by the National Science Foundation under Grant No. AST-2108168, AST-2108169, AST2307126, and AST-2407526.

This paper also made use of the Whole Sky Database (wsdb) created by Sergey Koposov and maintained at the Institute of Astronomy, Cambridge by Sergey Koposov, Vasily Belokurov, and Wyn Evans with financial support from the Science \& Technology Facilities Council (STFC) and the European Research Council (ERC).


This research has made use of the VizieR catalogue access tool, CDS, Strasbourg, France (\url{https://cds.u-strasbg.fr}). The original description of the VizieR service was published in \citet{VizieR2000}.

\end{acknowledgments}

\clearpage
\bibliography{bibliography.bib}{}
\bibliographystyle{aasjournal}

\appendix

\section{Data table} \label{appendix:data}

In Table \ref{tab:data_in_appendix}, we present the 1100 stars with good-quality data (see Section \ref{sec:data}) used for membership evaluation. The $P$ column shows the membership probability computed with the mixture modeling approach in this work (Section \ref{subsec:gmm}), while other quantities are taken from the \sfive DR2 catalog.

\renewcommand{\arraystretch}{1.0}
\setlength{\tabcolsep}{0.5em}
\begin{table*}[ht!]
\centering
\caption{\sfive DR2 data around the Crater~2 galaxy and stream footprint
}
\label{tab:data_in_appendix}
\begin{tabular}{ccccccccccc}
\hline
\textit{Gaia} DR3 \texttt{source\_id} & $\alpha$ & $\delta$ & $\mu_\alpha \cos{\delta}$ & $\mu_\delta$ & \vhel & $e_\vhel$ & \feh & $e_{\feh}$ & $P$ & $S/N$ \\
& (deg) & (deg) & (${\rm mas}\,{\rm yr}^{-1}$) & (${\rm mas}\,{\rm yr}^{-1}$) & ($\kms$) & & \\
\hline
        3544474297064516096 & 172.30925 & $-$21.31763 & $-$0.064 & $\phantom{-}$0.234 & $\phantom{-}$17.64$\phantom{0}$ & 2.94 & $-$2.35 & 0.25 & 0.00 & 5.9 \\
        3544502648144931968 & 172.34935 & $-$20.92886 & $-$0.471 & $-$0.603 & $\phantom{-}$288.63 & 2.49 & $-$1.71 & 0.16 & 0.00 & 8.5 \\
        3541353230230311552 & 172.52798 & $-$21.76984 & $\phantom{-}$0.354 & $-$0.69 & $-$1.86$\phantom{00}$ & 4.65 & $-$0.69 & 0.35 & 0.00 & 3.6 \\
        3544591979168107392 & 172.55615 & $-$20.38967 & $-$0.210 & $-$0.167 & $\phantom{-}$47.17$\phantom{0}$ & 3.09 & $-$0.78 & 0.33 & 0.00 & 5.2 \\
        3544973926316400128 & 172.58195 & $-$20.21502 & $\phantom{-}$0.064 & $-$1.107 & $-$1.51$\phantom{00}$ & 6.06 & $-$1.36 & 0.94 & 0.00 & 4.8 \\ 
        3541452907830980736 & 172.60148 & $-$21.54574 & $-$0.102 & $-$0.002 & $\phantom{-}$286.68 & 1.49 & $-$2.08 & 0.18 & 0.00 & 9.6 \\ 
        3544510615308138496 & 172.64702 & $-$20.95267 & $-$0.271 & $-$0.083 & $\phantom{-}$129.16 & 4.85 & $-$2.17 & 0.33 & 0.92 & 5.3 \\ 
        3541455145508547456 & 172.67600 & $-$21.46399 & $-$0.505 & $-$0.547 & $\phantom{-}$49.30$\phantom{0}$ & 7.33 & $-$2.47 & 0.66 & 0.00 & 4.2 \\ 
        3544512749906827904 & 172.69912 & $-$20.87969 & $\phantom{-}$1.208 & $-$0.435 & $\phantom{-}$12.03$\phantom{0}$ & 4.98 & $-$1.42 & 0.65 & 0.00 & 3.0 \\
\hline
\end{tabular}
\begin{flushleft}
The first 10 rows of this table are shown here to exemplify its format and content.
\end{flushleft}
\end{table*}



\end{document}